\documentclass[showpacs,preprintnumbers,amsmath,amssymb]{revtex4}
\usepackage{amsmath}
\usepackage{txfonts}

\usepackage{graphicx}
\usepackage{rotating}
\usepackage{dcolumn}
\usepackage{bm}

\begin{document}


\title{Bosonization, Singularity Analysis, Nonlocal Symmetry Reductions and Exact Solutions of Supersymmetric KdV Equation\footnote{Corresponding author: S. Y. Lou; Email address: sylou@sei.ecnu.edu.cn and lousenyue@nbu.edu.cn}}
\author{Xiao Nan Gao$^{1}$, S. Y. Lou$^{2,3}$ and Xiao Yan Tang$^{1}$}
\affiliation{$^{1}$Department of Physics, Shanghai Jiao Tong
University, Shanghai, 200240, China\\
$^{2}$Shanghai Key Laboratory of Trustworthy Computing, East China Normal University, Shanghai 200062, China\\
$^{3}$Faculty of Science, Ningbo University, Ningbo,
315211, China}

\date{\today}

\begin{abstract}
 Assuming that there exist at least two fermionic parameters, the classical $\mathcal {N}=1$ supersymmetric Korteweg-de Vries (SKdV) system can be transformed to some coupled bosonic systems. The boson fields in the bosonized SKdV (BSKdV) systems are defined on even Grassmann algebra. Due to the intrusion of other Grassmann parameters, the BSKdV systems are different from the usual non-suppersymmetric integrable systems, and many more abundant solution structures can be unearthed. With the help of the singularity analysis, the Painlev\'e property of the BSKdV system is proved and a B\"acklund transformation (BT) is found. The BT related nonlocal symmetry, we call it as residual symmetry, is used to find symmetry reduction solutions of the BSKdV system. Hinted from the symmetry reduction solutions, a more generalized but much simpler method is established to find exact solutions of the BSKdV and then the SKdV systems, which actually can be applied to any fermionic systems.
\end{abstract}
\pacs{11.30.Pb, 02.30.Ik, 11.10.Lm, 12.60.Jv, 04.65.+e, 02.30.Jr}

\maketitle

\section{Introduction}
In quantum field theory, the bosonization
approach is one of the powerful methods which simplifies the
procedure to treat complex fermionic fields \cite{qf}.
It is difficult to find a proper bosonization
procedure for both quantum and classical supersymmetric integrable
models though the supersymmetric quantum mechanical problems can be successfully bosonized \cite{4a}. To treat the integrable systems with fermions such as the super integrable systems \cite{super}, supersymmetric integrable systems \cite{SS} and pure integrable fermionic systems \cite{fer}
is much more complicated than to study the integrable pure bosonic
systems. Therefore, it is significant if one can establish a proper
bosonization procedure to treat the supersymmetric systems even if
in the classical level.

In our previous Letter \cite{Boson},
a bosonization approach with $N$ fermionic parameters to deal with
SKdV system is developed such that the SKdV system can be solved by the usual KdV equation
together with several linear differential equations without
fermionic variables. Especially, some types of exact
supersymmetric extensions of any solutions of the usual KdV equation
can be obtained straightforwardly through the exact solutions of the
KdV equation and the related symmetries.

The results of \cite{Boson} show us that for the SKdV equation there exist various kinds of localized excitations. In other words, in additional to the single supersymmetric traveling wave soliton solution (in the super
space-time $\{x,\ t,\ \theta\}$) known in literature \cite{BL},
there are infinitely many single traveling soliton extensions in the
usual space-time $\{x,\ t\}$.
The bosonization procedure and abundant properties of the soliton excitations of the classical
SKdV system reveals some open problems in both classical and quantum
theories. For instance, the fermionic fields take value on an infinite Grassmann algebra,
 that is on an algebra with infinite generators, however, in order to find some exact explicit solutions, we write down the solutions realized only on some finite dimensional Grassmann algebra \cite{Boson}. Hence, one of the important problems is how to obtain an extension to the case of infinite generators in the algebra. To modify this problem, in this paper, we extend the bosonization procedure of \cite{Boson} to a slightly generalized form by defining the bosonic fields on an infinite even Grassmann algebra $G_e$
 \begin{equation}\label{Ge}
 G_e=\left\{1,\ \prod_{i=1}^{2n}\zeta_i,\ n=1,\ 2,\ \cdots,\ \infty\right\},
 \end{equation}
 where $\zeta_i,\ i=1,\ 2,\ \cdots,\ \infty$ are usual Grassmann parameters with the anti-commutation property $\zeta_i\zeta_j=-\zeta_j\zeta_i$.
 Applying the new method, many more abundant nonlinear excitations of the SKdV system can be discovered.

Although the bosonic fields are still defined on an infinite even Grassmann algebra $G_e$, one essential advantage of the method is that it can effectively
avoid difficulties caused by intractable fermionic fields which are
anticommuting. The $\mathcal N =1$ supersymmetric versions of the
Korteweg-de Vries equation have been found for more than $20$ years
 \cite{2,3,4}, which are the beginning of the field of
supersymmetric integrable systems. The far-reaching significance
lies in not only mathematics, but also the applications in various
areas of modern theoretical physics especially in quantum field theory
and cosmology such as superstring theory  where it appears as a
basic part of the string worldsheet physics or the theory of
two-dimensional solvable lattice models, e.g., tricritical Ising
models \cite{McA,Kulish}. Therefore, investigating their
properties and searching for their exact solutions are of great
importance and interest.

For the integrable SKdV system in the sense of possessing a Lax
pair, many remarkable properties have been discovered, such as the
Painlev\'e property \cite{5}, the bi-Hamiltonian structures
\cite{6,7}, the Darboux transformation \cite{8}, the bilinear forms
\cite{9,10}, the B\"acklund transformation (BT) \cite{12}, the Lax
representation \cite{13} and the nonlocal conservation laws
\cite{14}. Some types of multisoliton solutions are also known for
the integrable SKdV system \cite{9,10,11,12,13,14}. However, because
anticommutative fermionic fields bring some difficulties in dealing
with supersymmetric equations, to get exact solutions of the
supersymmetric systems is, especially, much more difficult than pure bosonic systems.

In Sec. II of this paper, we simply review the results of \cite{Boson} and then slightly extend it by defining the bosonic fields from the usual c-number space to the infinite Grassmann subalgebra, $G_e$. The remaining sections are devoted to the study of a special BSKdV system (BSKdV-2) which is a special realization of the SKdV on a Grassmann-2 algebra $G_2\equiv \{1,\ \zeta_1,\ \zeta_2,\ \zeta_1\zeta_2\}$ in which the boson fields are still defined on the infinite even Grassmann algebra $G_e$. The similar results can be obtained for other special realizations of the SKdV (BSKdV-n) on a Grassmann-n algebra. In Sec. III, the Painlev\'e property and the B\"acklund transformation (BT) are studied by the standard singularity analysis. In Sec. IV, the infinitely many nonlocal symmetries are obtained starting from the BT related nonlocal symmetry, the residual symmetry (RS). The symmetry reductions related to both the local Lie point symmetries and the nonlocal RS are studied in Sec. V. According to the results of Sec. V, a more general but simpler method, the generalized tanh function expansion method, is proposed in Sec. VI to find exact solutions of the BSKdV-2 system. Some special explicit novel exact solutions of the BSKdV-2 and the SKdV are also investigated in Sec. VI. The last section includes a summary and some discussions.

\section{Review and Extension of the Bosonization of the SKdV equation}

The $\mathcal {N}=1$ supersymmetric version of the KdV equation,
\begin{eqnarray}\label{kdv}
u_t +6uu_x +u_{xxx} =0,
\end{eqnarray}
is established by extending the classical spacetime ($x, t$) to a
super-spacetime ($\theta, x, t$), where $\theta$ is a Grassmann
variable, and the field $u$ to a fermionic superfield
\begin{eqnarray}
\Phi(\theta, x, t) = \xi(x, t)+\theta u(x, t),
\end{eqnarray}
which leads to a nontrivial result \cite{3}
\begin{eqnarray}\label{Phi}
\Phi_t +3 (\mathcal {D} \Phi_x) \Phi +3 (\mathcal {D} \Phi) \Phi_x
+\Phi_{xxx} =0,
\end{eqnarray}
where $\mathcal {D} = \partial _{\theta} +\theta \partial _x$ is the
covariant derivative. The component version of Eq. \eqref{Phi} reads
\begin{subequations}\label{uxi}
\begin{equation}\label{u}
u_t +u_{xxx} -3\xi \xi_{xx} +6uu_x =0,
\end{equation}
\begin{equation}\label{xi}
\xi_t +\xi_{xxx} +3u_x \xi +3 u\xi_x =0,
\end{equation}
\end{subequations}
where $u$ and $\xi$ are bosonic and fermionic component fields,
respectively. Vanishing $\xi$ in Eq. \eqref{uxi}, only the usual
classical KdV equation remains.

Previous studies of the SKdV system were directly based on Eq.
\eqref{Phi} or \eqref{uxi}. In this paper, similar to \cite{Boson}, we are only concentrated on bosonization of the SKdV equations by expanding the supperfields with respect to the multi-fermionic parameters.

Firstly, we assume that for the solutions of the component fields $\xi$ and $u$ there exist at least two fermionic (Grassmann) parameters, say, $\zeta_1$ and  $\zeta_2$. Thus, we can expand the fields as
\begin{subequations}\label{uxi2}
\begin{equation}\label{u2}
\xi(x, t) =p \zeta_1 +q \zeta_2,
\end{equation}
\begin{equation}\label{xi2}
u(x, t) =u_0 +u_1 \zeta_1 \zeta_2 ,
\end{equation}
\end{subequations}
 where
the coefficients $p\equiv p(x,\ t)$, $q\equiv q(x,\ t)$, $u_0\equiv
u_0(x,\ t)$ and $u_1\equiv u_1(x,\ t)$ are four arbitrary functions with respect to the spacetime variables $x$ and
$t$, then the SKdV system \eqref{u}--\eqref{xi} is changed to the BSKdV-2 system
\begin{subequations}\label{bos}
\begin{equation}\label{bos1}
{u_0}_t +{u_0}_{xxx} +6u_0{u_0}_x=0,
\end{equation}
\begin{equation}\label{pt}
p_t +p_{xxx} +3 u_0p_x +3{u_0}_x p =0,
\end{equation}
\begin{equation}\label{qt}
q_t +q_{xxx} +3 u_0q_x +3{u_0}_x q =0,
\end{equation}
\begin{equation}\label{u1t}
{u_1}_t +{u_1}_{xxx} +6u_0{u_1}_x+6{u_0}_xu_1 =3(pq_{xx} -qp_{xx}).
\end{equation}
\end{subequations}
\bf Remark. \rm Though the forms of \eqref{uxi2} and \eqref{bos} are same as in \cite{Boson}, however, the connotation or sense of this paper is quite different from that of \cite{Boson}. In this paper, all the boson fields $p,\ q, \ u_0$ and $u_1$ are defined on not only the usual c-number algebra (the algebra constituted by the usual complex numbers) but also the infinite even Grassmann algebra $G_e$, while in \cite{Boson} all the fields take values only on the usual c-number algebra. In other words, various solutions of the SKdV equation will be lost if the bosoninc fields are not defined on $G_e$. As a simplest example, the BSKdV \eqref{bos1} possesses a solution
\begin{subequations}\label{J}
\begin{equation}
u_0=\vartheta_1\vartheta_2J,
\end{equation}
\begin{equation}
J_t+J_{xxx}=0,
\end{equation}
\end{subequations}
where $\vartheta_1$ and $\vartheta_2$ are Grassmann parameters. However, if $u_0$ is Grassmann parameter independent as in \cite{Boson}, the solution \eqref{J} will be lost.

Eq. \eqref{bos1} has the same form of the usual KdV equation but with depending on the Grassmann parameters. Eqs. \eqref{pt} and \eqref{qt} are linear homogeneous in
$p$ and $q$ respectively, and Eq. \eqref{u1t} is linear
nonhomogeneous in $u_1$.

Similarly, if there are at least $N\geq2$ fermionic
parameters $\zeta_{i}\ (i=1,2, \cdots, N)$ in a special solution of the SKdV system, we can expand the component fields $u$
and $\xi$ in the form
\begin{subequations}\label{uxiN}
\begin{equation}
u(x, t) =u_0 + \sum_{n=1}^{[\frac{N+1}{2}]} \sum_{1\leq i_{1}<
\cdots <i_{2n}\leq N} u_{i_{1}i_{2} \cdots i_{2n}} \zeta_{i_{1}}
\zeta_{i_{2}} \cdots \zeta_{i_{2n}} ,
\end{equation}
\begin{equation}
\xi(x, t) = \sum_{n=1}^{[\frac{N+1}{2}]} \sum_{1\leq i_{1}< \cdots
<i_{2n-1}\leq N} v_{i_{1}i_{2} \cdots i_{2n-1}} \zeta_{i_{1}}
\zeta_{i_{2}} \cdots \zeta_{i_{2n-1}},
\end{equation}
\end{subequations}
where the coefficients $u_{0}\equiv u_{0}(x,t)$, $u_{i_{1}i_{2}
\cdots i_{2n}}\equiv u_{i_{1}i_{2} \cdots i_{2n}}(x, t)\ (1 \leq
i_{1} <i_{2} < \cdots <i_{2n} \leq N)$ and $v_{i_{1}i_{2} \cdots
i_{2n-1}} \equiv v_{i_{1} i_{2} \cdots i_{2n-1}}(x, t)\ (1 \leq
i_{1} <i_{2} < \cdots <i_{2n-1} \leq N)$ are $2^N$
bosonic functions of classical spacetime variable $x$ and $t$ and also take values on the infinite even Grassmann algebra $G_e$.
Substituting Eq. \eqref{uxiN} into the SKdV model \eqref{uxi}, we
obtain the following bosonic system of $2^N$ equations, the BSKdV-N system,
\begin{subequations}\label{Eqn}
\begin{eqnarray}\label{Eqn-kdv}
{u_0}_t +{u_0}_{xxx} +6u_0{u_0}_x =0,
\end{eqnarray}
\begin{eqnarray}
L_o v_{i_1 i_2 \cdots i_{2n-1}} &=& \left\{\begin{array}{ll} 0
& \textrm{for $n=1$}\\
-3\sum\limits_{W_1} (-1)^{\tau(j_1,j_2, \cdots ,j_{2n-1})}
\left[u_{i_{j_1}i_{j_2} \cdots i_{j_{2l}}} v_{i_{j_{2l+1}}i_{j_{2l+2}}
\cdots i_{j_{2n-1}}}\right]_x & \textrm{for $n=2, 3, \cdots,
\left[\frac{N+1}{2}\right]$}
\end{array} \right. ,\label{Eqn-o}\\
L_e u_{i_1 i_2 \cdots i_{2n}} &=& \left\{\begin{array}{ll}
3\sum\limits_{W_2}(-1)^{\tau(j_1,j_2)} \left[v_{i_{j_{1}}}
(v_{i_{j_{2}}})_x\right]_x,
&\textrm{for $n =1$} \\
3\sum\limits_{W_2}(-1)^{\tau(j_1,j_2, \cdots ,j_{2n})}
\left[v_{i_{j_{1}}i_{j_{2}} \cdots i_{j_{2l-1}}}
(v_{i_{j_{2l}}i_{j_{2m+1}} \cdots i_{j_{2n}}})_x\right]_x \\
-3\sum\limits_{W_3} (-1)^{\tau(j_1,j_2, \cdots ,j_{2n})}
\left[u_{i_{j_1}i_{j_2} \cdots i_{j_{2l}}} u_{i_{j_{2l+1}}i_{j_{2l+2}}
\cdots i_{j_{2n}}}\right]_x     &\textrm{for $n=2, 3, \cdots,
\left[\frac{N}{2}\right]$}, \end{array} \right., \label{Eqn-e}
\end{eqnarray}
\end{subequations}
where
\begin{displaymath}
\tau(j_1,j_2, \cdots ,j_{N}) =\left\{\begin{array}{ll} 0 &
\textrm{for $j_1,\ j_2,\ \cdots,\ j_{N}$ is even permutation}
\\ 1 & \textrm{for $j_1,\ j_2,\ \cdots,\ j_{N}$ is odd permutation}
\end{array}\right. ,
\end{displaymath}
\begin{eqnarray}
W_1 &=&\left\{(j_1,j_2, \ldots j_{2n-1}) |1\leq j_1<j_2< \cdots < j_{2l}
\leq 2n-1, 1\leq j_{2l+1}<j_{2l+2}<\cdots < j_{2n-1} \leq 2n-1, \right.\nonumber\\
&&\left. 1\leq l\leq n-1, j_{h_1}\neq j_{h_2}(h_1\neq h_2)\right\}, \nonumber\\
W_2 &=& \left\{(j_1,j_2, \ldots j_{2n}) |1\leq j_1<j_2< \cdots < j_{2l-1}
\leq 2n, 1\leq j_{2l}<j_{2l+1}<\cdots < j_{2n}\leq 2n,\right. \nonumber\\
&&\left.1\leq l\leq n, j_{h_1}\neq j_{h_2}(h_1\neq h_2)\right\}, \nonumber\\
W_3 &=& \left\{(j_1,j_2, \ldots j_{2n}) |1\leq j_1<j_2< \cdots < j_{2l}
\leq 2n, 1\leq j_{2l+1}<j_{2l+2}<\cdots < j_{2n}\leq 2n,\right. \nonumber\\
&&\left. 1\leq l\leq n-1, j_{h_1}\neq j_{h_2}(h_1\neq h_2)\right\}, \nonumber
\end{eqnarray}
and two operators read
\begin{eqnarray}
L_e(u_0) &=&\partial_{t} +\partial_{xxx} +6u_0\partial_{x}
+6{u_0}_{x},\nonumber\\
L_o(u_0) &=&\partial_{t} +\partial_{xxx} +3u_0\partial_{x} +3{u_0}_{x}.
\nonumber
\end{eqnarray}
It is noted that the $N=2$ case, though the forms of \eqref{uxiN} and \eqref{Eqn} are same as those in \cite{Boson}, however, the connotation or sense here is different because all the boson fields $u_{0}$, $u_{i_{1}i_{2}
\cdots i_{2n}}$ and $v_{i_{1}i_{2} \cdots
i_{2n-1}}$ are defined on $G_e$ algebra while the boson fields of \cite{Boson} are defined only on the usual c-number algebra.

In this paper, we just concentrate on the BSKdB-2 system. To study the integrability and exact solutions of the BSKdV-2 system \eqref{bos}, it is convenient to study its singularity.

\section{Singularity Analysis and B\"acklund Transformations of the BSKdV-2 system.}

Because of the difficulty to study the nonlinear systems, there is no method to find general solutions of any nontrivial nonlinear system except for C-integrable systems which can be directly transformed to linear ones or can be solved by direct integration. To find Laurent series solutions is an effective way to get the final general solution in linear case. In nonlinear case, the similar method is called singularity analysis or the Painlev\'e test. By means of the Painlev\'e analysis, various integrable properties such as the B\"acklund transformations, Lax pairs, infinitely many symmetries, bilinear forms, and so on, can be easily found if the studied model possesses Painlev\'e property, i.e., it is Painlev\'e integrable. In this section, we study the Painlev\'e property (PP) and the B\"acklund transformation of the BSKdV-2.

A nonlinear model is called Painlev\'e integrable, i.e., possessing Painlev\'e property, if all the movable singularities of its solutions are only poles. To prove the PP of the BSKdV-2 system, we expand the Boson fields $u_0,\ p,\ q$ and $u_1$ as
\begin{eqnarray}
u_0=\sum_{j=0}^{\infty}u_{0j}\phi^{j-\alpha_1},\
p=\sum_{j=0}^{\infty}p_{j}\phi^{j-\alpha_2},\
q=\sum_{j=0}^{\infty}q_{j}\phi^{j-\alpha_3},\
u_1=\sum_{j=0}^{\infty}u_{1j}\phi^{j-\alpha_4}.\label{pp1}
\end{eqnarray}
BSKdV-2 system \eqref{bos} possesses PP implies: (i) The constants $\alpha_1,\ \alpha_2,\ \alpha_3,$ and $\alpha_4$ are all positive integers such that all the non-pole singularities, algebraic and logarithmic branch points, are ruled out; (ii) There are twelve arbitrary functions in the series expansion \eqref{pp1} because four partial differential equations in \eqref{bos} are third order. Furthermore, the function $\phi$ should be arbitrary such that ALL the possible singularities are included.

After finishing the detailed calculations, it is not difficult to find that the BSKdV-2 is really Painlev\'e integrable because \eqref{pp1} possesses the form
\begin{eqnarray}
u_0=\sum_{j=0}^{\infty}u_{0j}\phi^{j-2},\
p=\sum_{j=0}^{\infty}p_{j}\phi^{j-2},\
q=\sum_{j=0}^{\infty}q_{j}\phi^{j-2},\
u_1=\sum_{j=0}^{\infty}u_{1j}\phi^{j-3},\label{pp2}
\end{eqnarray}
where
\begin{eqnarray}
u_{00}&=&-2\phi_x^2,\ u_{01}=2\phi_{xx},\  u_{02}=\frac16\phi_x^{-2}\big(3\phi_{xx}^2
-\phi_t\phi_x-4\phi_x\phi_{xxx}\big),\nonumber\\
p_1&=&-\big(p_0\phi_x^{-1}\big)_x,\ p_2=\frac1{12}\phi_x^{-4}\big(p_0\phi_t\phi_x-8p_0\phi_x\phi_{xxx}
+21p_0\phi_{xx}^2-18p_{0x}\phi_x\phi_{xx}+6\phi_x^2p_{0xx}\big)\nonumber\\
q_1&=&-\big(q_0\phi_x^{-1}\big)_x,\ q_2=\frac1{12}\phi_x^{-4}\big(q_0\phi_t\phi_x-8q_0\phi_x\phi_{xxx}
+21q_0\phi_{xx}^2-18q_{0x}\phi_x\phi_{xx}+6\phi_x^2q_{0xx}\big)\nonumber\\
u_{11}&=&\frac12{\phi_x}^2\big(3u_{10}\phi_{xx}-2\phi_xu_{10x}\big),\nonumber\\
u_{12}&=&-\frac12\phi_x^4\big(2u_{10}\phi_x\phi_{xxx}+4u_{10}\phi_x\phi_{xx}
-\phi_x^2u_{10xx}-6u_{10}\phi_{xx}^2\big),\label{pp3}
\end{eqnarray}
and other functions $u_{0j},$ $ p_j,$ $ q_j$ and $u_{1j}$ are all determined by twelve arbitrary functions $\phi,$ $ p_0,$ $ q_0,$ $ u_{10},$ $ u_{04},$ $ p_{4},$ $ q_4,$ $ p_5,$ $ q_5,$ $ u_{15},$ $ u_{06},$ and $u_{17}$.

Now, using the standard truncated Painlev\'e expansion
\begin{eqnarray}
u_0&=&-2\phi_x^2\phi^{-2}+2\phi_{xx}\phi^{-1}+u_2,\nonumber\\
p&=&p_0\phi^{-2}-\big(p_0\phi_x^{-1}\big)_x\phi^{-1}+p_2,\nonumber\\
q&=&q_0\phi^{-2}-\big(q_0\phi_x^{-1}\big)_x\phi^{-1}+q_2,\nonumber\\
u_1&=&u_{10}\phi^{-3}+\frac12{\phi_x}^2\big(3u_{10}\phi_{xx}
-2\phi_xu_{10x}\big)\phi^{-2}-\frac12\phi_x^4\big(2u_{10}\phi_x\phi_{xxx}+4u_{10}\phi_x\phi_{xx}
-\phi_x^2u_{10xx}-6u_{10}\phi_{xx}^2\big)\phi^{-1}+u_{13},\label{BT1}
\end{eqnarray}
we have the following B\"acklund theorem:\\
\bf Theorem 1 (\it BT theorem). \rm If the fields $\phi,\ p_0,\ q_0$ and $u_{10}$ are the solutions of the following Schwarzian BSKdV-2 system,
\begin{eqnarray}
&&\phi_t+\phi_{xxx}-\frac3{2\phi_x}\phi_{xx}^2+\lambda\phi_x=0,\nonumber\\
&&p_{0t}+p_{0xxx}+\frac12\lambda p_{0x}+\frac{9 p_0}{\phi_x^{2}}\phi_{xxx}\phi_{xx}
-\frac{33}{2\phi_x^{3}}p_0\phi_{xx}^3+\frac{\lambda}{\phi_x} p_0\phi_{xx}-\frac9{2\phi_x}p_{0x}\phi_{xxx}-\frac6{\phi_x}p_{0xx}\phi_{xx}
+\frac{63}{4\phi_x^{2}}p_{0x}\phi_{xx}^2=0,\nonumber\\
&&q_{0t}+q_{0xxx}+\frac12\lambda q_{0x}+\frac{9q_0}{\phi_x^{2}}\phi_{xxx}\phi_{xx}
-\frac{33}{2\phi_x^{3}}q_0\phi_{xx}^3+\frac{\lambda q_0}{\phi_x}\phi_{xx}-\frac9{2\phi_x}q_{0x}\phi_{xxx}
-\frac{6}{\phi_x}q_{0xx}\phi_{xx}
+\frac{63}{4\phi_x^2}q_{0x}\phi_{xx}^2=0,\nonumber\\
&&u_{10t}+u_{10xxx}-\frac{9}{\phi_x}u_{10xx}\phi_{xx}
-\frac{u_{10x}}{\phi_x}\left({6}
\phi_{xxx}-\lambda\phi_x \right)-\frac{3}{\phi_x}(q_{0x}p_{0xx}-p_{0x}q_{0xx})
-\frac{6}{\phi_x}\phi_{xx}(p_0q_{0x}-q_0p_{0x})_x\nonumber\\
&&\quad +\frac{63}{2\phi_x^{2}}u_{10x}\phi_{xx}^2
+\frac{18}{\phi_x}u_{10}\phi_{xx}\phi_{xxx}
+\frac{p_0q_{0x}-q_0p_{0x}}{4\phi_x^3}\big(18\phi_{xxx}\phi_x-2\lambda\phi_x^2 +{33}\phi_{xx}^2\big)
-\frac{42u_{10}}{\phi_x}\phi_{xx}^3=0, \label{SBSKdV2}
\end{eqnarray}
then \eqref{BT1} is a BT between the solutions $\{u_0,\ p,\ q,\ u_1\}$ and $\{u_{2},\ p_2,\ q_2,\ u_{13}\}$  while the latter solution is related to $\{\phi,\ p_0,\ q_0,\ u_{10}\}$ by
\begin{eqnarray}\label{BTu2}
u_2&=&\frac12\frac{\phi_{xx}^2}{\phi_x^{2}}-\frac12\frac{\phi_{xxx}}{\phi_x}
+\frac{\lambda}6 ,\nonumber\\
p_2&=&-\frac{3p_0\phi_{xxx}}{4\phi_x^{3}}+\frac{15p_0\phi_{xx}^2}{8\phi_x^{4}}
-\frac{p_0}{12\phi_x^{2}}-\frac{3p_{0x}\phi_{xx}}{2\phi_x^{3}}
+\frac12p_{0xx}\phi_x^2,\nonumber\\
q_2&=&-\frac{3q_0\phi_{xxx}}{4\phi_x^{3}}+\frac{15q_0\phi_{xx}^2}{8\phi_x^{4}}
-\frac{q_0}{12\phi_x^{2}}-\frac{3q_{0x}\phi_{xx}}{2\phi_x^{3}}
+\frac12q_{0xx}\phi_x^2,\nonumber\\
u_{13}&=&\frac{\phi_t}{24\phi_x^5}\big(3q_0p_x-3p_0q_x
+\phi_xu_{10x}-3u_{10}\phi_{xx}\big)
+\frac{u_{10x}}{12\phi_x^5}\big(14\phi_x\phi_{xxx}-51\phi_{xx}^2\big)
-\frac{u_{10t}+4u_{10xxx}}{24\phi_x^3}\nonumber\\
&& +\frac{pq_{0t}-qp_{0t}+10u_{10x}\phi_{xx}
+u_{10}(\phi_t+3\phi_{xxx})_x}{8\phi_x^4} -\frac{u_{10}\phi_{xx}}{8\phi_x^6}\big(32\phi_x\phi_{xxx}-49\phi_{xx}^2\big).\label{seed}
\end{eqnarray}
It is obvious that the BT theorem 1 includes an auto-BT \eqref{BT1} and a nonauto-BT \eqref{BTu2}.

\section{Nonlocal symmetries of the BSKdV-2 from BT}
In the study of the nonlinear systems, the symmetry study is another powerful method. It is interesting that the truncated Painlev\'e expansion method can be used to find infinitely many nonlocal symmetries.

A symmetry of a nonlinear model is defined as a solution of its linearized system. For the BSKdV-2 system \eqref{bos}, its linearized system has the form
\begin{subequations}\label{boss}
\begin{equation}\label{boss1}
\sigma^{u_0}_t +\sigma^{u_0}_{xxx} +6\sigma^{u_0}{u_0}_x+6u_0\sigma^{u_0}_x=0,
\end{equation}
\begin{equation}\label{pts}
\sigma^p_t +\sigma^p_{xxx} +3 \sigma^{u_0}p_x +3\sigma^{u_0}_x p+3 u_0\sigma^p_x +3{u_0}_x \sigma^p =0,
\end{equation}
\begin{equation}\label{qts}
\sigma^q_t +\sigma^q_{xxx} +3 \sigma^{u_0}q_x +3\sigma^{u_0}_x q+3 u_0\sigma^q_x +3{u_0}_x \sigma^q =0,
\end{equation}
\begin{equation}\label{u1ts}
\sigma^{u_1}_t +\sigma^{u_1}_{xxx} +6(\sigma^{u_0}u_1)_x +6(u_0\sigma^{u_1})_x =3(\sigma^{p}q_{xx} -q\sigma^{p}_{xx})+3(p\sigma^{q}_{xx} -\sigma^{q}p_{xx}),
\end{equation}
\end{subequations}
which means \eqref{bos} is form invariant under the transformation
\begin{equation}\label{trans}
\{u_0,\ p,\ q,\ u_1\}\longrightarrow \{u_0,\ p,\ q,\ u_1\}+\epsilon \{\sigma^{u_0},\ \sigma^p,\ \sigma^q,\ \sigma^{u_1}\}
\end{equation}
with the infinitesimal parameter $\epsilon$.

For the integrable system there are infinitely many local and nonlocal symmetries. Recently, it is found that infinitely many nonlocal symmetries can be found from Darboux transformations \cite{LouHuDT,XP} or B\"acklund transformations \cite{XR}. From the truncated Painlev\'e expansion one can find not only BTs but also nonlocal symmetries.

For the BSKdV-2 system, a nonlocal symmetry
\begin{equation}\label{RS}
\sigma_{RS}=\left(\begin{array}{c} \sigma^{u_0}\cr \sigma^p \cr \sigma^q\cr \sigma^{u_1}\end{array}\right)=\left(\begin{array}{c} 2\phi_{xx}\cr -(p_0\phi_x^{-1})_x \cr -(q_0\phi_x^{-1})_x,\cr -\frac1{2\phi_x^4}\big(2u_{10}\phi_x\phi_{xxx}-6u_{10}\phi_{xx}^2
+4u_{10x}\phi_x\phi_{xx}-u_{10xx}\phi_x^2\big)\end{array}\right)
\end{equation}
can naturally read out from its truncated Painlev\'e expansion. In fact, the nonlocal symmetry \eqref{RS} is the residual of the truncated Painlev\'e expansion, i.e., BT \eqref{BT1} with respect to the singularity manifold $\phi$. Thus, we call this nonlocal symmetry as the residual symmetry (RS).

It is also interesting that RS \eqref{RS} is just the infinitesimal form of the BT \eqref{BT1}. To prove this conclusion, we have to solve the ``initial value problem"
\begin{eqnarray}\label{IV}
\frac{\mbox{d} {\bar{u}}_{0}(\epsilon)}{\mbox{d}\epsilon}
&=&2{\bar{\phi}}_{xx}(\epsilon),\quad {\bar{u}}_{0}(0)=u_0,\nonumber\\
\frac{\mbox{d} {\bar{p}}(\epsilon)}{\mbox{d}\epsilon}
&=&-\big[\bar{p}_0(\epsilon)\bar{\phi}_x(\epsilon)^{-1}\big]_x,\quad {\bar{p}}(0)=p,\nonumber\\
\frac{\mbox{d} {\bar{q}}(\epsilon)}{\mbox{d}\epsilon}
&=&-\big[\bar{q}_0(\epsilon)\bar{\phi}_x(\epsilon)^{-1}\big]_x,\quad {\bar{q}}(0)=q,\nonumber\\
\frac{\mbox{d} {\bar{u}}_{1}(\epsilon)}{\mbox{d}\epsilon}
&=&-\frac1{2\bar{\phi}(\epsilon)_x^4}\big[2\bar{u}_{10}(\epsilon)\bar{\phi}_x(\epsilon)
\bar{\phi}_{xxx}(\epsilon)-6\bar{u}_{10}(\epsilon)\bar{\phi}_{xx}(\epsilon)^2
+4\bar{u}_{10x}(\epsilon)\bar{\phi}_x(\epsilon)\bar{\phi}_{xx}(\epsilon)
-\bar{u}_{10xx}(\epsilon)\bar{\phi}_x(\epsilon)^2\big],\quad {\bar{u}}_{1}(0)=u_1.
\end{eqnarray}
It is clear that to solve the initial value problem \eqref{IV} is quite difficult due to the intrusion of the functions $\{\phi,\ p_0,\ q_0,\ u_{10}\}$ and their differentiations.

To solve the initial value problem \eqref{IV}, we have to prolong the BSKdV-2 system \eqref{bos} such that RS becomes a local Lie point symmetry for the prolonged system. Fortunately, the final result can be successfully obtained similar to the usual KdV equation and other integrable systems \cite{LouHuDT,XP,XR,NLS}:
\begin{subequations}\label{bose}
\begin{equation}\label{bose1}
{u_0}_t +{u_0}_{xxx} +6u_0{u_0}_x=0,
\end{equation}
\begin{equation}\label{pte}
p_t +p_{xxx} +3 u_0p_x +3{u_0}_x p =0,
\end{equation}
\begin{equation}\label{qte}
q_t +q_{xxx} +3 u_0q_x +3{u_0}_x q =0,
\end{equation}
\begin{equation}\label{u1te}
{u_1}_t +{u_1}_{xxx} +6u_0{u_1}_x+6{u_0}_xu_1 =3(pq_{xx} -qp_{xx}).
\end{equation}
\begin{equation}
\phi_t+\phi_{xxx}-\frac32\frac{\phi_{xx}^2}{\phi_x}+\lambda\phi_x=0,
\end{equation}
\begin{equation}
p_{0t}+p_{0xxx}+\frac12\lambda p_{0x}+9p_0\frac{\phi_{xxx}\phi_{xx}}{\phi_x^{2}}
-\frac{33p_0\phi_{xx}^3}{2\phi_x^{3}}+\lambda p_0\frac{\phi_{xx}}{\phi_x}-\frac92\frac{p_{0x}\phi_{xxx}}{\phi_x}
-6p_{0xx}\frac{\phi_{xx}}{\phi_x}
+\frac{63}4\frac{p_{0x}\phi_{xx}^2}{\phi_x^{2}}=0,
\end{equation}
\begin{equation}
q_{0t}+q_{0xxx}+\frac12\lambda q_{0x}+9q_0\frac{\phi_{xxx}\phi_{xx}}{\phi_x^{2}}
-\frac{33q_0\phi_{xx}^3}{2\phi_x^{3}}+\lambda q_0\frac{\phi_{xx}}{\phi_x}-\frac92\frac{q_{0x}\phi_{xxx}}{\phi_x}
-6q_{0xx}\frac{\phi_{xx}}{\phi_x}
+\frac{63}4\frac{q_{0x}\phi_{xx}^2}{\phi_x^{2}}=0,
\end{equation}
\begin{eqnarray}
&&u_{10t}+u_{10xxx}-\frac{9u_{10xx}\phi_{xx}}{\phi_x}
-\frac{u_{10x}}{\phi_x}\big(6
\phi_{xxx}-\lambda \phi_x \big)-\frac3{\phi_x}(q_{0x}p_{0xx}-p_{0x}q_{0xx})-\frac{6\phi_{xx}}{\phi_x}(p_0q_{0x}
-q_0p_{0x})_x\nonumber\\
&&\quad +\frac{63}{2\phi_x}u_{10x}\phi_{xx}^2
+\frac{18u_{10}}{\phi_x}\phi_{xx}\phi_{xxx}
+\frac{p_0q_{0x}-q_0p_{0x}}{4\phi_x^3}\big(18\phi_{xxx}\phi_x-2\lambda\phi_x^2 +{33}\phi_{xx}^2\big)
-\frac{42u_{10}\phi_{xx}^3}{\phi_x^{3}}=0,
\end{eqnarray}
\begin{equation}
\phi_x=\phi_1,
\end{equation}
\begin{equation}
\phi_{1x}=\phi_2,
\end{equation}
\begin{equation}
\phi_{2x}=\phi_3,
\end{equation}
\begin{equation}
u_{10x}=v_1,
\end{equation}
\begin{equation}
v_{1x}=v_2,
\end{equation}
\begin{equation}
p_{0x}=p_1,
\end{equation}
\begin{equation}
q_{0x}=q_1.
\end{equation}
\end{subequations}
Now it is not difficult to verify that the nonlocal symmetry, the RS symmetry \eqref{RS}, for the original BSKdV-2 system \eqref{bos} becomes a local Lie point symmetry for the prolonged system \eqref{bose}
\begin{equation}\label{RSL}
\sigma_{rs}=\left(\begin{array}{c} \sigma^{u_0}\cr \sigma^p \cr \sigma^q\cr \sigma^{u_1}\cr \sigma^{\phi}\cr \sigma^{p_0}\cr \sigma^{q_0}\cr \sigma^{u_{10}}\cr \sigma^{\phi_1}\cr \sigma^{\phi_2}\cr \sigma^{\phi_3}\cr \sigma^{p_1}\cr \sigma^{q_1}\cr \sigma^{v_1}\cr \sigma^{v_2}\end{array}\right)=\left(\begin{array}{c} 2\phi_2\cr p_0\phi_2\phi_1^{-2}-p_1\phi_1^{-1} \cr
q_0\phi_2\phi_1^{-2}-q_1\phi_1^{-1}\cr \frac1{2}\big(6u_{10}\phi_2^2 -2u_{10}\phi_1\phi_3
-4v_1\phi_1\phi_2+v_2\phi_1^2\big)\phi_1^{-4} \cr
-\phi^2 \cr -4p_0\phi \cr -4q_0\phi \cr -6u_{10}\phi \cr
-2\phi\phi_1 \cr -2\phi_1^2-2\phi\phi_2 \cr -6\phi_1\phi_2-2\phi\phi_3\cr
-4p_0\phi_1-4\phi p_1 \cr -4q_0\phi_1-4\phi q_1 \cr
-6u_{10}\phi_1-6\phi v_1\cr -6u_{10}\phi_2-12 v_1\phi_2-6\phi v_2
\end{array}\right).
\end{equation}
Correspondingly, the initial value problem \eqref{IV} is changed as
\begin{subequations}\label{IV1}
\begin{eqnarray}
\frac{\mbox{d} {\bar{u}}_{0}(\epsilon)}{\mbox{d}\epsilon}
&=&2{\bar{\phi}}_2(\epsilon),\quad {\bar{u}}_{0}(0)=u_0,
\end{eqnarray}
\begin{eqnarray}
\frac{\mbox{d} {\bar{p}}(\epsilon)}{\mbox{d}\epsilon}
&=&\bar{p}_0(\epsilon)\bar{\phi}_2(\epsilon)\bar{\phi}_1(\epsilon)^{-2}
-\bar{p}_1(\epsilon)\bar{\phi}_1(\epsilon)^{-1}
,\quad {\bar{p}}(0)=p,
\end{eqnarray}
\begin{eqnarray}
\frac{\mbox{d} {\bar{q}}(\epsilon)}{\mbox{d}\epsilon}
&=&\bar{q}_0(\epsilon)\bar{\phi}_2(\epsilon)\bar{\phi}_1(\epsilon)^{-2}
-\bar{q}_1(\epsilon)\bar{\phi}_1(\epsilon)^{-1}
,\quad {\bar{q}}(0)=q,
\end{eqnarray}
\begin{eqnarray}
\frac{\mbox{d} {\bar{u}}_{1}(\epsilon)}{\mbox{d}\epsilon}
&=&-\frac1{2\bar{\phi}_1(\epsilon)^4}\big[2\bar{u}_{10}(\epsilon)\bar{\phi}_1(\epsilon)
\bar{\phi}_3(\epsilon)-6\bar{u}_{10}(\epsilon)\bar{\phi}_2(\epsilon)^2
+4\bar{v}_{1}(\epsilon)\bar{\phi}_1(\epsilon)\bar{\phi}_2(\epsilon)
-\bar{v}_{2}(\epsilon)\bar{\phi}_1(\epsilon)^2\big],\quad {\bar{u}}_{1}(0)=u_1,
\end{eqnarray}
\begin{eqnarray}
\frac{\mbox{d} {\bar{\phi}}(\epsilon)}{\mbox{d}\epsilon}
&=& -\bar{\phi}(\epsilon)^2,\ \bar{\phi}(0)=\phi,
\end{eqnarray}
\begin{eqnarray}
\frac{\mbox{d} {\bar{p}_0}(\epsilon)}{\mbox{d}\epsilon}
&=& -4\bar{p}_0(\epsilon)\bar{\phi}(\epsilon), \ \bar{p}_0(0)=p_0,
\end{eqnarray}
\begin{eqnarray}
\frac{\mbox{d} {\bar{q}_0}(\epsilon)}{\mbox{d}\epsilon}
&=& -4\bar{q}_0(\epsilon)\bar{\phi}(\epsilon),\ \bar{q}_0(0)=q_0,
\end{eqnarray}
\begin{eqnarray}
\frac{\mbox{d} {\bar{u}_{10}}(\epsilon)}{\mbox{d}\epsilon}
&=& -6\bar{u}_{10}(\epsilon)\bar{\phi}(\epsilon),\ \bar{u}_{10}(0)=u_{10},
\end{eqnarray}
\begin{eqnarray}
\frac{\mbox{d} {\bar{\phi}_1}(\epsilon)}{\mbox{d}\epsilon}
&=&-2\bar{\phi}(\epsilon)\bar{\phi}_1(\epsilon),\ \bar{\phi}_1(0)=\phi_1,
\end{eqnarray}
\begin{eqnarray}
\frac{\mbox{d} {\bar{\phi}_2}(\epsilon)}{\mbox{d}\epsilon}
&=& -2\bar{\phi}_1(\epsilon)^2-2\bar{\phi}(\epsilon)\bar{\phi}_2(\epsilon), \ \bar{\phi}_2(0)=\phi_2,
\end{eqnarray}
\begin{eqnarray}
\frac{\mbox{d} {\bar{\phi}_3}(\epsilon)}{\mbox{d}\epsilon}
&=& -6\bar{\phi}_1(\epsilon)\bar{\phi}_2(\epsilon)-2\bar{\phi}(\epsilon)\bar{\phi}_3(\epsilon),
 \ \bar{\phi}_3(0)=\phi_3,
\end{eqnarray}
\begin{eqnarray}
\frac{\mbox{d} {\bar{p}_1}(\epsilon)}{\mbox{d}\epsilon}
&=&-4\bar{p}_0(\epsilon)\bar{\phi}_1(\epsilon)-4\bar{\phi}(\epsilon) \bar{p}_1(\epsilon),
 \ \bar{p}_1(0)=p_1,
\end{eqnarray}
\begin{eqnarray}
\frac{\mbox{d} {\bar{q}_1}(\epsilon)}{\mbox{d}\epsilon}
&=& -4\bar{q}_0(\epsilon)\bar{\phi}_1(\epsilon)-4\bar{\phi}(\epsilon) \bar{q}_1(\epsilon),
 \ \bar{q}_1(0)=q_1,
\end{eqnarray}
\begin{eqnarray}
\frac{\mbox{d} {\bar{v}_1}(\epsilon)}{\mbox{d}\epsilon}
&=&-6\bar{u}_{10}(\epsilon)\bar{\phi}_1(\epsilon)-6\bar{\phi}(\epsilon) \bar{v}_1(\epsilon),
 \ \bar{v}_1(0)=v_1,
\end{eqnarray}
\begin{eqnarray}
\frac{\mbox{d} {\bar{v}_2}(\epsilon)}{\mbox{d}\epsilon}
&=&-6\bar{u}_{10}(\epsilon)\bar{\phi}_2(\epsilon)-12 \bar{v}_1(\epsilon)\bar{\phi}_2(\epsilon)-6\bar{\phi}(\epsilon) \bar{v}_2(\epsilon),
 \ \bar{v}_2(0)=v_2.
\end{eqnarray}
\end{subequations}
The solution of the initial value problem \eqref{IV1} can be written as the following BT theorem for the enlarged BSKdV-2 system \eqref{bose}:\\
\bf Theorem 2 (\it equivalent BT theorem). \rm If $\{u_0,\ p,\ q,\ u_1,\ \phi,\ p_0,\ q_0,\ u_{10},\ \phi_1,\ \phi_2,\ \phi_3,\ p_1,\ q_1,\ v_1,\ v_2\}$ is a solution of the prolonged BSKdV-2 system \eqref{bose}, so is $\{\bar{u}_0,\ \bar{p},\ \bar{q},\ \bar{u}_1,\ \bar{\phi},\ \bar{p}_0,\ \bar{q}_0,\ \bar{u}_{10},\ \bar{\phi}_1,\ \bar{\phi}_2,\ \bar{\phi}_3,\ \bar{p}_1,\ \bar{q}_1,\ \bar{v}_1,\ \bar{v}_2\}$ with
\begin{subequations}\label{BTE}
\begin{eqnarray}
\bar{u}_0=u_0+\frac{2\epsilon\phi_2}{\epsilon\phi+1}-\frac{2\epsilon^2\phi_1^2}
{(\epsilon\phi+1)^2},\label{BTE1}
\end{eqnarray}
\begin{eqnarray}
\bar{p}=p-\frac{\epsilon p_0[\epsilon \phi_1^2+\phi_2(1+\epsilon\phi)]}{\phi_1^2(1+\epsilon\phi)^2}
+\frac{\epsilon p_1}{\phi_1(1+\epsilon\phi)},
\label{BTE2}
\end{eqnarray}
\begin{eqnarray}
\bar{q}=q-\frac{\epsilon q_0[\epsilon \phi_1^2+\phi_2(1+\epsilon\phi)]}{\phi_1^2(1+\epsilon\phi)^2}
+\frac{\epsilon q_1}{\phi_1(1+\epsilon\phi)},
\label{BTE3}
\end{eqnarray}
\begin{eqnarray}
\bar{u}_1=u_1+\epsilon u_{10}\left(\frac{3\phi_2^2-\phi_1\phi_3}
{\phi_1^4(1+\epsilon\phi)}+\frac{3\epsilon\phi_2}
{\phi_1^2(1+\epsilon\phi)^2}+\frac{\epsilon^2}
{(1+\epsilon\phi)^3}\right)-\frac{\epsilon v_1}{\phi_1(1+\epsilon\phi)}\left(\frac{2\phi_2}{\phi_1^2}
+\frac{\epsilon}{1+\epsilon\phi}\right)+\frac{\epsilon v_2}{2\phi_1^2(1+\epsilon\phi)},\label{BTE4}
\end{eqnarray}
\begin{eqnarray}
\bar{\phi}=\frac{\phi}{1+\epsilon\phi},
\end{eqnarray}
\begin{eqnarray}
\bar{p}_0=\frac{p_0}{(1+\epsilon\phi)^4},
\end{eqnarray}
\begin{eqnarray}
\bar{q}_0=\frac{q_0}{(1+\epsilon\phi)^4},
\end{eqnarray}
\begin{eqnarray}
\bar{u}_{10}=\frac{u_{10}}{(1+\epsilon\phi)^6},
\end{eqnarray}
\begin{eqnarray}
\bar{\phi}_1=\frac{\phi_1}{(1+\epsilon\phi)^2},
\end{eqnarray}
\begin{eqnarray}
\bar{\phi}_2=\frac{\phi_2}{(1+\epsilon\phi)^2}-\frac{2\phi_1^2\epsilon}
{(1+\epsilon\phi)^3},
\end{eqnarray}
\begin{eqnarray}
\bar{\phi}_3=\frac{\phi_3}{(1+\epsilon\phi)^2}-\frac{6\epsilon\phi_1\phi_2}
{(1+\epsilon\phi)^3}+\frac{6\epsilon^2\phi_1^3}
{(1+\epsilon\phi)^4}
\end{eqnarray}
\begin{eqnarray}
\bar{p}_1=\frac{p_1}
{(1+\epsilon\phi)^4}-\frac{4\epsilon\phi_1p_0}
{(1+\epsilon\phi)^5},
\end{eqnarray}
\begin{eqnarray}
\bar{q}_1=\frac{q_1}
{(1+\epsilon\phi)^4}-\frac{4\epsilon\phi_1q_0}
{(1+\epsilon\phi)^5},
\end{eqnarray}
\begin{eqnarray}
\bar{v}_1=\frac{v_1}
{(1+\epsilon\phi)^6}-\frac{6\epsilon\phi_1u_{10}}
{(1+\epsilon\phi)^7},
\end{eqnarray}
\begin{eqnarray}
\bar{v}_2=\frac{v_2}
{(1+\epsilon\phi)^6}-\frac{12\epsilon \phi_1 v_1}
{(1+\epsilon\phi)^7}-\frac{6\epsilon u_{10}[\epsilon\phi_2(1+\epsilon\phi)-7\epsilon\phi_1^2]}
{(1+\epsilon\phi)^8}.
\end{eqnarray}
\end{subequations}
For the original system \eqref{bos}, it is not difficult to see that the BT theorems 1 and 2 are equivalent based on the trivial fact that the singularity manifold equation \eqref{SBSKdV2} is form invariant under the transformation
$$1+\epsilon\phi \rightarrow \phi, \ (\epsilon\phi_1\rightarrow \phi_x,\ \epsilon\phi_2\rightarrow \phi_{xx}).$$

The BT equivalent theorem 2 shows us an interesting result that the nonlocal RS of the truncated Painlev\'e expansion is just the infinitesimal form of the BT.

Notice that the BSKdV-2 is not explicitly $\lambda$ dependent while the RS is explicitly $\lambda$ dependent, then infinitely many nonlocal symmetries can be obtained in two ways, respectively,
\begin{eqnarray}
\sigma_{RS,i}=\sigma_{RS}(\lambda_i), i=0,\ 1,\ 2,\ \cdots,
\end{eqnarray}
and
\begin{eqnarray}
\sigma_{RS,n}=\frac1{n!}\frac{\mbox{d}^n}{\mbox{d}\lambda^n}\sigma_{RS}, n=0,\ 1,\ 2,\ \cdots.
\end{eqnarray}
For the usual KdV system, the parameter $\lambda$ is just the spectral parameter, $\sigma_{RS}$ is equivalent to the square spectral function symmetry \cite{Loucom}, and the square spectral function symmetry form of $\sigma_{RS,i}$ can be used to find algebraic geometry solutions with higher order genus \cite{Cao}. $\sigma_{RS,n}$ for the usual KdV equation can be equivalently obtained by applying the inverse recursion $\Phi_{KdV}$ on $\sigma_{RS}$ \cite{neg}:
$$\sigma_{RS,n}(KdV) \sim \Phi_{KdV}^{-n} \sigma_{RS} \sim \Phi_{KdV}^{-n} \phi_{xx}. $$

\section{Symmetry reductions related to the nonlocal RS}
In Ref. \cite{Boson}, the symmetry reductions without nonlocal symmetry $\sigma_{RS}$ have been given for the fields defined on the usual c-number algebra but not on $G_e$. The symmetry reduction forms, if the boson fields are also defined on $G_e$, have the same forms as those in \cite{Boson}, except that all the reduction functions and parameters are defined on $G_e$. In this section, we are interested in studying the symmetry reductions of the BSKdV-2 system related to the nonlocal symmetry RS, $\sigma_{RS}$.

Similar to the last section, because of the nonlocal property of $\sigma_{RS}$, we have to study the symmetry reductions of the prolonged system \eqref{bose} since the RS has been localized to the Lie point symmetry \eqref{RSL}.

Using the standard Lie point symmetry approach to the prolonged system \eqref{bose}, we can find the general Lie point symmetry in the form
\begin{equation}\label{RSP}
\sigma_{pt}=\left(\begin{array}{c} \sigma^{u_0}\cr \sigma^p \cr \sigma^q\cr \sigma^{u_1}\cr \sigma^{\phi}\cr \sigma^{p_0}\cr \sigma^{q_0}\cr \sigma^{u_{10}}\cr \sigma^{\phi_1}\cr \sigma^{\phi_2}\cr \sigma^{\phi_3}\cr \sigma^{p_1}\cr \sigma^{q_1}\cr \sigma^{v_1}\cr \sigma^{v_2}\cr
\sigma^{\lambda}
\end{array}\right)=\left(\begin{array}{c}
(cx+x_0)u_{0x}+(3ct+t_0)u_{0t}+2cu_0+2C_0\phi_2 \cr
(cx+x_0)p_{x}+(3ct+t_0)p_{t}+c_2p+C_0(p_0\phi_2\phi_1^{-2}-p_1\phi_1^{-1}) \cr
(cx+x_0)q_{x}+(3ct+t_0)q_{t}+c_3q+C_0(q_0\phi_2\phi_1^{-2}-q_1\phi_1^{-1}) \cr
(cx+x_0)u_{1x}+(3ct+t_0)u_{1t}+(c_2+c_3-c)u_1
+\frac{C_0}{2}\big(6u_{10}\phi_2^2 -2u_{10}\phi_1\phi_3
-4v_1\phi_1\phi_2+v_2\phi_1^2\big)\phi_1^{-4} \cr
(cx+x_0)\phi_{x}+(3ct+t_0)\phi_{t}+(c_1-c)\phi+c_0-C_0\phi^2 \cr
(cx+x_0)p_{0x}+(3ct+t_0)p_{0t}+(c_2+2c_1-2c)p_0-4C_0p_0\phi \cr (cx+x_0)q_{0x}+(3ct+t_0)q_{0t}+(c_3+2c_1-2c)q_0-4C_0q_0\phi \cr
(cx+x_0)u_{10x}+(3ct+t_0)u_{10t}+(c_3+c_2-4c+3c_1)u_{10}-6C_0u_{10}\phi \cr
(cx+x_0)\phi_{1x}+(3ct+t_0)\phi_{1t}+c_1\phi_1-2C_0\phi\phi_1 \cr
(cx+x_0)\phi_{2x}+(3ct+t_0)\phi_{2t}+(c_1+c)\phi_2-2C_0\phi_1^2-2C_0\phi\phi_2 \cr
(cx+x_0)\phi_{3x}+(3ct+t_0)\phi_{3t}+(c_1+2c)\phi_3
-6C_0\phi_1\phi_2-2C_0\phi\phi_3\cr
(cx+x_0)p_{1x}+(3ct+t_0)p_{1t}+(c_2+2c_1-c)p_1-4C_0p_0\phi_1-4C_0\phi p_1 \cr
(cx+x_0)q_{1x}+(3ct+t_0)q_{1t}+(c_3+2c_1-c)q_1-4C_0q_0\phi_1-4C_0\phi q_1 \cr
(cx+x_0)v_{1x}+(3ct+t_0)v_{1t}+(c_2+c_3+3c_1-3c)v_1-6C_0u_{10}\phi_1-6C_0\phi v_1\cr
(cx+x_0)v_{2x}+(3ct+t_0)v_{2t}+(c_2+c_3+3c_1-2c)v_2-6C_0(u_{10}\phi_2+2 v_1\phi_2+\phi v_2)\cr
2\lambda c
\end{array}\right),
\end{equation}
which includes the space-time scaling invariance ($c$-part of \eqref{RSP}), space, time and the fields $\{\phi,\ p_1,\ q_1,\ v_1,\ v_2\}$ translations ($x_0$, $t_0$ and $\{c_0,\ C_1,\ C_2,\ C_3,\ C_4\}$ parts, respectively), RS ($C_0$ part) and the scaling transformations of the fields
$\phi,\ p$ and $q$ ($c_1,\ c_2$ and $c_3$ parts, respectively). Because we are concentrated on the RS related symmetry reductions in this paper, we fix $C_0=1$ without loss of generality for $C_0\neq 0$. We also fix the trivial potential fields $\{p_1,\ q_1,\ v_1,\ v_2\}$ translations by taking  $C_1=C_2=C_3=C_4=0$ for simplicity. The last component of $\sigma_{pt}$ shown in \eqref{RSP} indicates that to guarantee the space-time scaling invariance, the spectral parameter $\lambda$ should also have a scaling like the field $u_0$.

To find symmetry reductions, i.e., to find group invariant solutions, implies to find solutions of both
\begin{equation}\label{s=0}
\sigma_{pt}=0
\end{equation}
and the prolonged system \eqref{bose}. To solve the group invariant condition \eqref{s=0}, we have to discuss two nontrivial cases for $c\neq 0$ and $c=0$.\\
\bf Case I $c\neq 0$. \rm In this case the general solution of \eqref{s=0} with $C_1=C_2=C_3=C_4=C_0-1=0$ has the form
\begin{subequations}\label{g1}
\begin{eqnarray}
\lambda=0,\ \xi=\frac{a_1x-3b_1x_0}{3T},\ T\equiv(a_1t-b_1t_0),
\end{eqnarray}
\begin{eqnarray}
u_0=T^{-\frac23}\left[U_0+\frac2{a_1}\Psi_2\tanh(W)
-\frac2{a_1^2}\Psi_1^2\tanh^2(W)\right],\ W\equiv b_1\ln\left(-\frac{a_1t}{b_1}+t_0\right)+\Psi(\xi),
\end{eqnarray}
\begin{eqnarray}
p=T^{\frac{c_2b_1}{a_1}}\left[P+\frac{P_0\Psi_2-P_1\Psi_1}{a_1\Psi_1^2}\tanh(W)
-\frac1{a_1^2}P_0\tanh^2(W)\right],
\end{eqnarray}
\begin{eqnarray}
q=T^{\frac{c_3b_1}{a_1}}\left[Q+\frac{Q_0\Psi_2-Q_1\Psi_1}{a_1\Psi_1^2}\tanh(W)
-\frac1{a_1^2}P_0\tanh^2(W)\right],
\end{eqnarray}
\begin{eqnarray}
u_1=T^{\frac{b_1(c_2+c_3)}{a_1}+\frac13}\left[U_1-\frac{2U_{10}(\Psi_1\Psi_3
-3\Psi_2^2)-4V_1\Psi_1\Psi_2-V_2\Psi_1^2}{2a_1\Psi_1^4}\tanh(W)
+\frac{3U_{10}\Psi_2-2V_1\Psi_1}{a_1^2\Psi_1^2}\tanh^2(W)
+\frac{U_{10}}{a_1^3}\tanh^3(W)\right],
\end{eqnarray}
\begin{eqnarray}
\Psi=a_0+a_1\tanh(W),
\end{eqnarray}
\begin{eqnarray}
p_0=T^{\frac{c_2b_1}{a_1}}P_0\mbox{sech}^4(W),
\end{eqnarray}
\begin{eqnarray}
q_0=T^{\frac{c_3b_1}{a_1}}Q_0\mbox{sech}^4(W),
\end{eqnarray}
\begin{eqnarray}
u_{10}=T^{\frac{(c_2+c_3)b_1}{a_1}+\frac13}U_{10}\mbox{sech}^6(W),
\end{eqnarray}
\begin{eqnarray}
\Psi_1=T^{-\frac13}\Psi_{1}\mbox{sech}^2(W),
\end{eqnarray}
\begin{eqnarray}
\Psi_2=T^{-\frac23}\mbox{sech}^2(W)\left[\Psi_2-\frac2{a_1}\Psi_1^2\tanh(W)\right],
\end{eqnarray}
\begin{eqnarray}
\Psi_3=\frac1T\mbox{sech}^2(W)\left[\Psi_3-\frac6{a_1}\Psi_1\Psi_2\tanh(W)
+\frac6{a_1^2}\Psi_1^3\tanh^2(W)\right],
\end{eqnarray}
\begin{eqnarray}
p_1=T^{\frac{c_2b_1}{a_1}-\frac13}\mbox{sech}^4(W)\left[P_1
-\frac4{a_1}\Psi_1P_0\tanh(W)\right],
\end{eqnarray}
\begin{eqnarray}
q_1=T^{\frac{c_3b_1}{a_1}-\frac13}\mbox{sech}^4(W)\left[Q_1
-\frac4{a_1}\Psi_1Q_0\tanh(W)\right],
\end{eqnarray}
\begin{eqnarray}
v_1=T^{\frac{b_1}{a_1}(c_2+c_3)}\mbox{sech}^6(W)\left[V_1
-\frac6{a_1}\Psi_1u_{01}\tanh(W)\right],
\end{eqnarray}
\begin{eqnarray}
v_2=T^{\frac{b_1}{a_1}(c_2+c_3)-\frac13}\mbox{sech}^6(W)\left[V_2
-\frac6{a_1}(\Psi_2u_{01}+2\Psi_1V_1)\tanh(W)
+\frac{42}{a_1^2}U_{10}\Psi_1^2\tanh^2(W)\right],
\end{eqnarray}
\end{subequations}
where $U_0,\ P,\ Q,\ U_1,\ \Psi,\ P_0,\ Q_0,\ U_{10},\ \Psi_1,\ \Psi_2,\ \Psi_3,\ P_1,\ Q_1,\ V_1$ and $V_2$ are all group invariant functions of $\xi$ and the constants $c,\ c_0$ and $c_1$ in \eqref{RSP} have been re-notated as $$c = -\frac{3a_1}{b_1},\ c_0 = -a_0^2+a_1^2,\ c_1 = \frac{6b_1a_0-a_1}{3b_1}$$
for notation simplicity.

Substituting the group invariant solution \eqref{g1} into the prolonged system \eqref{bose}, it is straightforward to get the symmetry reduction equations for the group invariant functions $U_0,\ P,\ Q,\ U_1,\ \Psi,\ P_0,\ Q_0,\ U_{10},\ \Psi_1,\ \Psi_2,\ \Psi_3,\ P_1,\ Q_1,\ V_1$ and $V_2$:
\begin{subequations}\label{red1}
\begin{eqnarray}
U_0=\frac1{2\Psi_1^2}\left(a_1^2b_1\Psi_1-\Psi_2^2-\xi\Psi_1^2\right),
\end{eqnarray}
\begin{eqnarray}
\Psi_3=\frac1{2\Psi_1}\left(3\Psi_2^2-2a_1^2b_1\Psi_1-2\xi\Psi_1^2\right),
\end{eqnarray}
\begin{eqnarray}
\Psi'=\frac3{a_1^2}\Psi_1,
\end{eqnarray}
\begin{eqnarray}
P'_{0}=\frac3{a_1}P_1,
\end{eqnarray}
\begin{eqnarray}
Q'_{0}=\frac3{a_1}Q_1,
\end{eqnarray}
\begin{eqnarray}
U'_{10}=\frac3{a_1}V_1,
\end{eqnarray}
\begin{eqnarray}
\Psi'_{1}=\frac3{a_1}\Psi_2, \label{red1phi2}
\end{eqnarray}
\begin{eqnarray}
V'_{1}=\frac3{a_1^2}(6U_{10}\Psi_1^2+a_1^2V_2),
\end{eqnarray}
\begin{eqnarray}
\Psi'_{2}=\frac3{2a_1^3\Psi_1}(4\Psi_1^4
+2a_1^2\xi\Psi_1^2+3a_1^2\Psi_2^2-2a_1^4b_1\Psi_1), \label{red1phi1}
\end{eqnarray}
\begin{eqnarray}
P'_{1}=\frac32\left(\frac{8}{a_1^3}\Psi_1^2+\frac{3\xi}{a_1}
-\frac{3a_1b_1}{\Psi_1}-\frac{3\Psi_2^2}{a_1\Psi_1^2}\right)P_0
+\frac6{a_1}P\Psi_1^2+\frac{9P_1\Psi_2}{a_1\Psi_1},
\end{eqnarray}
\begin{eqnarray}
Q'_{1}=\frac32\left(\frac{8}{a_1^3}\Psi_1^2+\frac{3\xi}{a_1}
-\frac{3a_1b_1}{\Psi_1}-\frac{3\Psi_2^2}{a_1\Psi_1^2}\right)Q_0
+\frac6{a_1}Q\Psi_1^2+\frac{9Q_1\Psi_2}{a_1\Psi_1},
\end{eqnarray}
\begin{eqnarray}
P'=\frac{3\Psi_2}{\Psi_1}P+\frac{3}{2a_1}\left(\frac2{a_1^2}
+\frac{\xi}{\Psi_1^2}\right)P_1-\frac3{4a_1}
\left(\frac{4\Psi_2}{a_1^2\Psi_1}+\frac{a_1+2b_1c_2}{\Psi_1^2}
+\frac{3\xi\Psi_2}{\Psi_1^3}\right)P_0,
\end{eqnarray}
\begin{eqnarray}
Q'=\frac{3\Psi_2}{\Psi_1}Q+\frac{3}{2a_1}\left(\frac2{a_1^2}
+\frac{\xi}{\Psi_1^2}\right)Q_1-\frac3{4a_1}
\left(\frac{4\Psi_2}{a_1^2\Psi_1}+\frac{a_1+2b_1c_2}{\Psi_1^2}
+\frac{3\xi\Psi_2}{\Psi_1^3}\right)Q_0,
\end{eqnarray}
\begin{eqnarray}
V_2&=&4\left(2\Psi_1Q_0-\frac{Q_1\Psi_1^2}{\Psi_2}\right)P
+4\left(\frac{P_1\Psi_1^2}{\Psi_2}-2P_0\Psi_1\right)Q
+\left(\frac{2\xi \Psi_1-b_1a_1^2}{\Psi_1}-\frac{5\Psi_2^2}{\Psi_1^2}
+\frac{\Psi_1}{2\Psi_2}(a_1+c_3b_1+c_2b_1)\right)U_{10}\nonumber\\
&&
+\left(\frac{5\Psi_2}{\Psi_1}-\frac{a_1^2b_1}{2\Psi_2}\right)V_1
+\left(\frac{a_1^2b_1P_1}{2\Psi_1\Psi_2}-\frac{P_1\Psi_2}{\Psi_1^2}
-\frac{b_1P_0}{2\Psi_2}(c_2-c_3)\right)Q_0+\left(\frac{\Psi_2}{\Psi_1^2}
-\frac{a_1^2b_1}{2\Psi_1\Psi_2}\right)P_0Q_1-4\frac{\Psi_1^4U_1}{\Psi_2},
\end{eqnarray}
\begin{eqnarray}
U'_{1}&=&-\frac{3V_2}{4a_1\Psi_1^3}\left[\frac{\Psi_2^2}{\Psi_1^2}+\xi
+\frac{4\Psi_1^3-a_1^4b_1}{2a_1^2\Psi_1}\right]
+\frac{3V_1}{4a_1\Psi_1^3}\left\{\frac{5\Psi_2^3}{\Psi_1^3}
+\left[\frac{5\xi}{\Psi_1}
+\frac{32\Psi_1^3-11a_1^4b_1}{4a_1^2\Psi_1^2}\right]\Psi_2
+\frac12(a_1+c_3 b_1+c_2 b_1)\right\}\nonumber\\
&&
+\frac{3U_{10}}{2a_1\Psi_1^3}\left\{\xi^2-\frac{5\Psi_2^4}{2\Psi_1^4}
-\left[\frac{3\xi}{2\Psi_1^2}
+\frac{6\Psi_1^3-a_1^4 b_1}{2a_1^2\Psi_1^3}\right]\Psi_2^2
-\frac{3(a_1+c_3 b_1+c_2 b_1)\Psi_2}{8\Psi_1}
+\frac{\xi(4\Psi_1^3-3 a_1^4 b_1)}{2a_1^2\Psi_1}
-\frac{b_1(5\Psi_1^3-a_1^4b_1)}{2\Psi_1^2}\right\}\nonumber\\
&&+\frac{3P_0}{\Psi_1^2}\left\{\frac{7b_1 (c_3-c_2)Q_0\Psi_2}{16 a_1\Psi_1^3}
+\left[\frac{\Psi_2^3}{4a_1\Psi_1^5}
+\left(\frac{\xi}{4a_1\Psi_1^3}-\frac{3a_1b_1}{16\Psi_1^4}\right)\Psi_2
+\frac{c_3 b_1+3 c_2 b_1+2 a_1}{8\Psi_1^2 a_1}\right] Q_1-\frac{2Q\Psi_2^2}{a_1\Psi_1^2}
-\frac{Q\xi}{2a_1}+\frac{Qa_1b_1}{4\Psi_1}\right\}\nonumber\\
&&+\frac{3 Q_0}{\Psi_1^2}\left\{-\left[\frac{\Psi_2^3}{4a_1\Psi_1^5}
+\left(\frac{\xi}{4 a_1\Psi_1^3}-\frac{3a_1 b_1}{16\Psi_1^4}\right)\Psi_2
+\frac{3c_3b_1+c_2 b_1+2 a_1}{8a_1\Psi_1^2}\right]P_1
+\frac{2 P\Psi_2^2}{\Psi_1^2 a_1}+\frac{P\xi}{2a_1}
-\frac{ P a_1 b_1}{4\Psi_1}\right\}\nonumber\\
&&+\frac{3Q\Psi_2 P_1}{\Psi_1^3 a_1}
-\frac{3P\Psi_2 Q_1}{\Psi_1^3 a_1}+\frac{3\Psi2 U_1}{2\Psi_1 a_1},
\end{eqnarray}
\end{subequations}
where, and in the latter of the paper, the primes `$'$' on the functions (with only one independent variable) denote derivatives with respect to the corresponding independent variable.

Eliminating $\Psi_2$ in \eqref{red1phi1} by using \eqref{red1phi2}, we get a single second order ordinary differential equation for the field $\Psi_1$
\begin{eqnarray}
\Psi''_{1}=\frac9{2a_1^4\Psi_1}\left(4\Psi_1^4
+2a_1^2\xi\Psi_1^2+\frac{a_1^4}3\Psi'^2_{1}-2a_1^4b_1\Psi_1\right), \label{red1psi1}
\end{eqnarray}
which is an equivalent form of the Painlev\'e II equation. Thus, the group invariant solution is an interaction solution among a soliton and a Painlev\'e II wave. \\
\bf Case II $c=0$. \rm The general solution of \eqref{s=0} with $c=C_1=C_2=C_3=C_4=C_0-1=0$ has the form
\begin{subequations}\label{g2}
\begin{eqnarray}
\eta=x-Ct,\ (x_0\equiv C t_0,\ t_0\equiv -a_1/b_1,\ c_1\equiv 2\Psi_0,\ c_0\equiv a_1^2-\Psi_0^2),
\end{eqnarray}
\begin{eqnarray}
u_0=U_0-\frac2{a_1}\Psi_2\tanh(W)
-\frac2{a_1^2}\Psi_1^2\tanh^2(W),\ W\equiv b_1 t+\Psi,
\end{eqnarray}
\begin{eqnarray}
p=e^{\frac{c_2b_1}{a_1}t}\left[P-\frac{P_1\Psi_1 -P_0\Psi_2}{a_1\Psi_1^2}\tanh(W)
+\frac1{a_1^2}P_0\tanh^2(W)\right],
\end{eqnarray}
\begin{eqnarray}
q=e^{\frac{c_3b_1}{a_1}t}\left[Q-\frac{Q_1\Psi_1 -Q_0\Psi_2}{a_1\Psi_1^2}\tanh(W)
+\frac1{a_1^2}Q_0\tanh^2(W)\right],
\end{eqnarray}
\begin{eqnarray}
u_1=e^{\frac{b_1(c_2+c_3)t}{a_1}}\left[U_1+\frac{2U_{10}(
3\Psi_2^2-\Psi_1\Psi_3)-4V_1\Psi_1\Psi_2+V_2\Psi_1^2}{2a_1\Psi_1^4}\tanh(W)
+\frac{3U_{10}\Psi_2-2V_1\Psi_1}{2a_1^2\Psi_1^2}\tanh^2(W)
-\frac{U_{10}}{a_1^3}\tanh^3(W)\right],
\end{eqnarray}
\begin{eqnarray}
\Psi=\Psi_0+a_1\tanh(W),
\end{eqnarray}
\begin{eqnarray}
p_0=e^{\frac{c_2b_1t}{a_1}}P_0\mbox{sech}^4(W),
\end{eqnarray}
\begin{eqnarray}
q_0=e^{\frac{c_3b_1t}{a_1}}Q_0\mbox{sech}^4(W),
\end{eqnarray}
\begin{eqnarray}
u_{10}=e^{\frac{(c_2+c_3)b_1t}{a_1}}U_{10}\mbox{sech}^6(W),
\end{eqnarray}
\begin{eqnarray}
\Psi_1=\Psi_{1}\mbox{sech}^2(W),
\end{eqnarray}
\begin{eqnarray}
\Psi_2=\mbox{sech}^2(W)\left[\Psi_2+\frac2{a_1}\Psi_1^2\tanh(W)\right],
\end{eqnarray}
\begin{eqnarray}
\Psi_3=\mbox{sech}^2(W)\left[\Psi_3-\frac6{a_1}\Psi_1\Psi_2\tanh(W)
+\frac6{a_1^2}\Psi_1^3\tanh^2(W)\right],
\end{eqnarray}
\begin{eqnarray}
p_1=e^{\frac{c_2b_1t}{a_1}}\mbox{sech}^4(W)\left[P_1
-\frac4{a_1}\Psi_1P_0\tanh(W)\right],
\end{eqnarray}
\begin{eqnarray}
q_1=e^{\frac{c_3b_1t}{a_1}}\mbox{sech}^4(W)\left[Q_1
-\frac4{a_1}\Psi_1Q_0\tanh(W)\right],
\end{eqnarray}
\begin{eqnarray}
v_1=e^{\frac{b_1(c_2+c_3)t}{a_1}}\mbox{sech}^6(W)\left[V_1
-\frac6{a_1}\Psi_1u_{01}\tanh(W)\right],
\end{eqnarray}
\begin{eqnarray}
v_2=e^{\frac{b_1(c_2+c_3)t}{a_1}}\mbox{sech}^6(W)\left[V_2
-\frac6{a_1}(\Psi_2u_{01}+2\Psi_1V_1)\tanh(W)
+\frac{42}{a_1^2}U_{10}\Psi_1^2\tanh^2(W)\right],
\end{eqnarray}
\end{subequations}
where $U_0,\ P,\ Q,\ U_1,\ \Psi,\ P_0,\ Q_0,\ U_{10},\ \Psi_1,\ \Psi_2,\ \Psi_3,\ P_1,\ Q_1,\ V_1$ and $V_2$ are all group invariant solution of $\eta$.

Substituting the group invariant solution \eqref{g2} into the prolonged system \eqref{bose}, we get the second symmetry reduction:
\begin{subequations}\label{red2}
\begin{eqnarray}
U_0=-\frac C2 +\frac{2\lambda}3+ \frac{a_1b_1}{2\Psi_1}-\frac{\Psi_2^2}{2\Psi_1^2},
\end{eqnarray}
\begin{eqnarray}
\Psi_3=\frac{\Psi_2^2}{2\Psi_1}+\left(\frac{\lambda}3-2U_0\right)\Psi_1,
\end{eqnarray}
\begin{eqnarray}
\Psi'=\frac{\Psi_1}{a_1},
\end{eqnarray}
\begin{eqnarray}
P'_{0}=P_1,
\end{eqnarray}
\begin{eqnarray}
Q'_{0}=Q_1,
\end{eqnarray}
\begin{eqnarray}
U'_{10}=V_1,
\end{eqnarray}
\begin{eqnarray}
\Psi'_{1}=\Psi_2, \label{red2phi2}
\end{eqnarray}
\begin{eqnarray}
V'_{1}=\frac1{a_1^2}(6U_{10}\Psi_1^2+a_1^2V_2),
\end{eqnarray}
\begin{eqnarray}
\Psi'_{2}=\left(C-\lambda\right)\Psi_1-a_1b_1+\frac32\frac{\Psi_2^2}{\Psi_1}
+\frac{2\Psi_1^3}{a_1^2}, \label{red2phi1}
\end{eqnarray}
\begin{eqnarray}
P'_{1}=\frac2{a_1^2}(2P_0+a_1^2P)\Psi_1^2-\frac13P_0(9U_0-2\lambda)
+\frac3{\Psi_1}\Psi_2P_1-\frac3{\Psi_1^2}\Psi_2^2P_0,
\end{eqnarray}
\begin{eqnarray}
Q'_{1}=\frac2{a_1^2}(2Q_0+a_1^2Q)\Psi_1^2-\frac13Q_0(9U_0-2\lambda)
+\frac3{\Psi_1}\Psi_2Q_1-\frac3{\Psi_1^2}\Psi_2^2Q_0,
\end{eqnarray}
\begin{eqnarray}
P'=\frac{\Psi_2}{\Psi_1}\left[P+P_0\left(\frac{24U_0}{\Psi_1^2}
-\frac1{a_1^2}+\frac{45C-62\lambda}{4\Psi_1^2}
-\frac{12a_1b_1}{\Psi_1^2}\right)\right]-\frac{b_1c_2P_0}{2a_1\Psi_1^2}
+\left(\frac1{a_1^2}+\frac{3C-2\lambda}{6\Psi_1^2}\right)P_1
+\frac{12\Psi_2^3P_0}{\Psi_1^5},
\end{eqnarray}
\begin{eqnarray}
Q'=\frac{\Psi_2}{\Psi_1}\left[Q+Q_0\left(\frac{24U_0}{\Psi_1^2}
-\frac1{a_1^2}+\frac{45C-62\lambda}{4\Psi_1^2}
-\frac{12a_1b_1}{\Psi_1^2}\right)\right]-\frac{b_1c_2Q_0}{2a_1\Psi_1^2}
+\left(\frac1{a_1^2}+\frac{3C-2\lambda}{6\Psi_1^2}\right)Q_1
+\frac{12\Psi_2^3Q_0}{\Psi_1^5},
\end{eqnarray}
\begin{eqnarray}
V_2&=&8\Psi_1(PQ_0-QP_0)-\frac{45\Psi_2^2U_{10}}{2\Psi_1^2}
+\left(\frac{64\lambda}3-\frac{31C}2-35U_0
-\frac{33a_1b_1}{2\Psi_1}\right)U_{10}+\frac1{\Psi_2}\left\{\left[2Q_0U_0+
4Q\Psi_1^2+\left(C-\frac{2\lambda}3
+\frac{a_1b_1}{2\Psi_1}\right)Q_0\right]P_1\right.\nonumber\\
&&
-\left.\left[2P_0U_0+
4P\Psi_1^2+\left(C-\frac{2\lambda}3
+\frac{a_1b_1}{2\Psi_1}\right)P_0\right]Q_1-\left[10\Psi_1U_0
+5\left(C-\frac{4\lambda}3\right)\Psi_1+\frac92a_1b_1\right]V_1
\right.\nonumber\\
&&\left.-\frac{c_2+c_3}{2a_1}b_1\Psi_1U_{10}+\frac{c_2-c_3}{2a_1}b_1Q_0P_0-4U_1
\Psi_1^4\right\},
\end{eqnarray}
\begin{eqnarray}
U'_{1}&=&\frac {V_2}{\Psi_1^3}\left(\frac{U_0}2-\frac{a_1 b_1}{8\Psi_1}-\frac{\Psi_1^2}{2a_1^2}
-\frac{\lambda}{12}\right)
-\left\{\frac{405 \Psi_2^3}{16\Psi_1^6}+\frac{\Psi_2}{\Psi_1^4} \left(\frac{425}8 U_0-\frac{205}6 \lambda
-\frac{2\Psi_1^2}{a_1^2}+\frac{405}{16}  C-\frac{207}{8\Psi_1} a_1 b_1\right)
-\frac{b_1(c2+c3)}{8\Psi_1^3 a_1}\right\}V_1\nonumber\\
&&+\left\{\frac{87 Q_0 \Psi_2^3}{8\Psi_1^7}
+\left[\left(\frac{89 U_0}{4}-\frac{175 b_1 a_1}{16\Psi_1} +\frac{29}{8} (3 C-4 \lambda)\right) \frac{Q_0}{\Psi_1^5}
+\frac{Q}{\Psi_1^3}\right] \Psi_2-\frac{b_1 Q_0(3 c_3+c_2)}{8 a_1\Psi_1^4}\right\}P_1
+\left\{-\frac{87\Psi_2^3 P_0}{8\Psi_1^7}\right.\nonumber\\
&&\left.
+\left[\left(-\frac{89 U_0}{4\Psi_1^5}+\frac{175 b_1 a_1}{16\Psi_1^6}-\frac{29(3 C-4 \lambda)}{8\Psi_1^5}\right) P_0
-\frac{P}{\Psi_1^3}\right] \Psi_2+\frac{b_1 P_0(3 c_2+c_3)}{8a_1\Psi_1^4}\right\}Q_1+\frac{153 U_{10} \Psi_2^4}{16\Psi_1^7}
\nonumber\\
&&
+\left[\frac{U_1}{2\Psi_1}-\frac{3 b_1(c_2+c_3)}{16 a_1\Psi_1^4} U_{10}
-\frac{7 b_1(c_2-c_3)}{16\Psi_1^5 a_1} P_0 Q_0\right]\Psi_2+\left\{-\frac{173 U_0^2}{4\Psi_1^3}
+\left[\frac{3}{a_1^2\Psi_1}-\frac{501 C-674 \lambda}{12\Psi_1^3}+\frac{171 b_1 a_1}{4\Psi_1^4}\right] U_0\right.\nonumber\\
&&\left.
+\frac{5 C-6 \lambda}{2 a_1^2\Psi_1}-\frac{11 b_1}{4a_1\Psi_1^2}
-\frac{319\lambda^2}{18\Psi_1^3}-\frac{153 C^2}{16\Psi_1^3} +\frac{313\lambda C}{12\Psi_1^3}
+\frac{3 a_1 b_1 }{8\Psi_1^4} (53 C-72 \lambda)-\frac{165 a1^2 b_1^2}{16\Psi_1^5} \right\}U_{10}\nonumber\\
&&
-\frac1{12 \Psi_1^3} (Q_0 P-P_0 Q)(18\Psi_1 C-28 \Psi1\lambda-21 b_1 a_1 +48\Psi_1 U_0).
\end{eqnarray}
\end{subequations}
This reduction system can be readily solved because \eqref{red2phi1} is only an autonomous second order ordinary differential equation by using \eqref{red2phi2},
\begin{eqnarray}
\psi''_{1}=\left(C-\lambda\right)\psi_1-b_1
+\frac32\frac{\psi'^2_{1}}{\psi_1}
+2\psi_1^3,\ \psi_1\equiv a_1\Psi_1. \label{Red2phi1}
\end{eqnarray}
The general solution of \eqref{Red2phi1} can be expressed by the following incomplete elliptic integral
\begin{eqnarray}
\int \frac{\mbox{d}\psi_1}{\sqrt{4\psi_1^4+C_0\psi_1^3
-2(C-\lambda)\psi_1^2+b_1\psi_1}}=\eta-\eta_0. \label{Phi1}
\end{eqnarray}
We do not discuss more about the solution of \eqref{red2} here but come back in the next section.
\section{Generalized Tanh Function Expansion Method of BSKdV-2 system}
\leftline{\bf{A. Tanh Function Expansion Method}}
\rm
According to the first four equations of the symmetry reductions of \eqref{red1} and \eqref{red2}, we can develop a simple method, the tanh function expansion method, to get more general solutions of the BSKdV-2 system. The usual tanh function expansion method is used only to find single traveling soliton or solitary wave solutions of nonlinear systems. The result of the last section implies that the tanh function expansion method can be extended to get many more exact solutions.

The generalized expansion solution for the both reductions \eqref{red1} and \eqref{red2} has the form
\begin{subequations}\label{tanh}
\begin{eqnarray}
u_0=u_{00}+u_{01}\tanh(w)+u_{02}\tanh^2(w),
\end{eqnarray}
\begin{eqnarray}
p=p_{0}+p_{1}\tanh(w)+p_{2}\tanh^2(w),
\end{eqnarray}
\begin{eqnarray}
q=q_{0}+q_{1}\tanh(w)+q_{2}\tanh^2(w),
\end{eqnarray}
\begin{eqnarray}
u_1=u_{10}+u_{11}\tanh(w)+u_{12}\tanh^2(w)+u_{13}\tanh^3(w),
\end{eqnarray}
\end{subequations}
where $w,\ u_{00},\ u_{01},\ u_{02},\ u_{10},\ u_{11},\ u_{12},\ u_{13},\ p_{0},\ p_{1},\ p_{2},\ q_{0},\ q_{1},$ and $ q_{2} $ are functions of $\{x,\ t\}$ and should be determined later.

After some direct calculations by substituting \eqref{tanh} into the BSKdV-2 system \eqref{bos}, we can prove the following nonauto-B\"acklund transformation theorem:\\
\bf Theorem 3 (\it Nonauto-BT theorem). \rm
If $\{w,\ f,\ g,\ h\}$ is a solution of
\begin{subequations}\label{bosw}
\begin{eqnarray}
w_t+w_{xxx}-\frac32w_{xx}^2w_x^{-1}-2w_x^3+\lambda w_x=0,
\end{eqnarray}
\begin{eqnarray}
f_t+f_{xxx}-\lambda f_x -\frac32 f_x w_t w_x^{-1}=0,
\end{eqnarray}
\begin{eqnarray}
g_t+g_{xxx}-\lambda g_x -\frac32 g_x w_t w_x^{-1}=0,
\end{eqnarray}
\begin{eqnarray}
h_t+h_{xxx}-2\lambda h_x + \lambda_1+(\lambda f-3f_{xx})g_x-(\lambda g-3g_{xx})f_x-\frac32(2h_x+gf_{x}-fg_{x})w_tw_x^{-1}=0,
\end{eqnarray}
\end{subequations}
then $\{u_0,\ p,\ q,\ u_1\}$ with
\begin{subequations}\label{solufg}
\begin{eqnarray}
u_0=-2w_x^2\tanh^2(w)+2w_{xx}\tanh(w)+\frac12w_tw_x^{-1}
-\frac12w_{xx}^2w_x^{-2}+\frac23\lambda,
\end{eqnarray}
\begin{eqnarray}
p=fw_x^2\tanh^2(w)-(fw_x)_x\tanh(w)+\frac12f_x w_{xx}w_x^{-1}+\frac f4 w_{xx}^2w_x^{-2}-\frac f4w_tw_x^{-1}-\frac f3\lambda+\frac12f_{xx},
\end{eqnarray}
\begin{eqnarray}
q=gw_x^2\tanh^2(w)-(gw_x)_x\tanh(w)+\frac12g_x w_{xx}w_x^{-1}+\frac g4 w_{xx}^2w_x^{-2}-\frac g4w_tw_x^{-1}-\frac g3\lambda+\frac12g_{xx},
\end{eqnarray}
\begin{eqnarray}
u_1&=&hw_x^3\tanh^3(w)-\frac{w_x}2(2h_xw_x+3hw_{xx})\tanh^2(w)
+\frac14\left(4w_{xx}h_x+3hw_{xx}^2w_x^{-1}-2h\lambda w_x-2hw_t+2w_xh_{xx}\right)\tanh(w)\nonumber\\
&&+\frac{\lambda_1}6-\frac{w_t}{8w_x^2}\big[w_x(h_x+gf_x-fg_x)-h w_{xx}\big]-\frac16f_x\left(\lambda g-3g_{xx}\right)
+\frac16g_x\left(\lambda f-3f_{xx}\right)+\frac18(h_t+fg_t-gf_t)\nonumber\\
&& +\frac1{8w_x}(hw_{xt}-2h_{xx}w_{xx}+2\lambda hw_{xx})-\frac{w_{xx}^2}{8w_x^3}(hw_{xx}+2h_xw_w),
\end{eqnarray}
\end{subequations}
is a solution of the BSKdV-2 system \eqref{bos}. \\
\bf{B. Exact solutions}\\
\rm At first glance, to find solutions of \eqref{bosw} is still difficult. However, it is interesting that we can find that some nontrivial solutions of the BSKdV-2 from some quite trivial solutions of \eqref{bosw}. Here are some interesting examples. \\
\bf Example 1 \it The Single soliton. \rm A quite trivial straight-line solution of \eqref{bosw} has the form
\begin{subequations}\label{line}
\begin{eqnarray}
w=kx+\omega t+x_0,\ p=k_1x+\omega_1t+x_1,\ q=k_2x+\omega_2t+x_2,\ h=k_3x+\omega_3t+x_3,\
\end{eqnarray}
\begin{eqnarray}
\omega=k(2k^2-\lambda),\ \omega_1=\frac{k_1}2(6k^2-\lambda),\ \omega_2=\frac{k_2}2(6k^2-\lambda),\ \omega_3=\frac{2k_3+k_1x_2-k_2x_1}2(6k^2-\lambda)-\lambda_1,
\end{eqnarray}
\end{subequations}
where all the free constants $k,\ k_1,\ k_2,\ k_3,\ \lambda,\ \lambda_1,\ x_0,\ x_1,\ x_2$ and $x_3$ can all be defined on not only the usual c-number algebra but also the even Grassmann algebra $G_e$. Substituting the line solution \eqref{line} into the nonauto-BT theorem yields the following soliton solution of the BSKdV and then the SKdV system
\begin{subequations}\label{line}
\begin{eqnarray}
u_0=2k^2\mbox{sech}^2(w)-k^2-\frac{\lambda}6,\ w\equiv kx+(2k^3-k\lambda)t+x_0,\ h\equiv k_3x+\omega_3t+x_3,
\end{eqnarray}
\begin{eqnarray}
p=f\left(k^2\tanh^2(w)-\frac{k^2}2+\frac{7\lambda}{12}\right)-kk_1\tanh(w), \ f\equiv k_1x+\omega_1t+x_1,
\end{eqnarray}
\begin{eqnarray}
q=g\left(k^2\tanh^2(w)-\frac{k^2}2+\frac{7\lambda}{12}\right)-kk_2\tanh(w), \ g\equiv k_2x+\omega_2t+x_2,
\end{eqnarray}
\begin{eqnarray}
u_1=-hk^3\mbox{sech}^2(w)\tanh(w)+\frac12k_3k^2(2\mbox{sech}^2(w)-1)
+\frac{(30k^2-\lambda)(k_2f-k_1g)}{48}+\frac{\lambda_1}{24}
-\frac{(6k^2-\lambda)(k_2x_1-k_1x_2)}{16}.
\end{eqnarray}
\end{subequations}
Though the soliton solution \eqref{line} is a traveling wave in the $\{x,\ t\}$ space-time for the boson field $u_0$, it is not a traveling wave for other boson fields $p,\ q$ and $u_1$, and then the superfiled $\Phi$ of SKdV is not a traveling wave except for the case of $f,\ g,$ and $h$ being constants, i.e., $k_1=k_2=k_3=\lambda_1=0$.

This example reveals that the nonauto-BT theorem is an straightening transformation of the BSKdV-2 which straightens the single soliton to a straight-line solution. \\
\bf Example 2 \em Painlev\'e II extensions\rm.
It is known that for the usual KdV system, there exists a Painlev\'e II reduction if one uses the scaling symmetry. In the same way, applying the scaling invariance (and space time translations) to the field $w$, we have the scaling group invariant solution
\begin{eqnarray}\label{ew2}
w=c_1\ln (t-t_0)+\Psi(\xi),\qquad \xi\equiv \frac{x-x_0}{(t-t_0)^{1/3}},
\end{eqnarray}
with the equivalent Painlev\'e II reduction
\begin{eqnarray}\label{re2}
\Psi''_{1}=\frac32\Psi'^2_{1}\Psi_1^{-1}+2\Psi_1^3+\frac13\xi\Psi_1-c_1,
\end{eqnarray}
where $c_1,\ x_0$ and $t_0$ are constants defined on $G_e$.

To find the corresponding solution for other fields $f,\ g$ and $h$, we have to solve
\begin{subequations}\label{fgh1}
\begin{eqnarray}\label{eqf1}
(t-t_0)f_{1t}+f_{1\xi\xi\xi}+\frac16\Psi_1^{-1}f_{1\xi}(\xi\Psi_1-9c_1)=0, f\equiv f_1(\xi,\ t), \ g\equiv g_1(\xi,\ t),
\end{eqnarray}
\begin{eqnarray}\label{eqg1}
(t-t_0)g_{1t}+g_{1\xi\xi\xi}+\frac16\Psi_1^{-1}g_{1\xi}(\xi\Psi_1-9c_1)=0,
\ h\equiv h_1(\xi,\ t)-\lambda_1(t-t_0),
\end{eqnarray}
\begin{eqnarray}\label{equ1}
(t-t_0)h_{1t}+h_{1\xi\xi\xi}+\frac13\Psi_1^{-1}h_{1\xi}(2\xi\Psi_1-9c_1)
+\frac12f_{1\xi}[6g_{1\xi\xi}+g_1\Psi_1^{-1}(\xi\Psi_1-3c_1)]
-\frac12g_{1\xi}[6f_{1\xi\xi}+f_1\Psi_1^{-1}(\xi\Psi_1-3c_1)]=0.
\end{eqnarray}
\end{subequations}
A special solution of \eqref{fgh1} has the form
\begin{subequations}\label{rfgh1}
\begin{eqnarray}
f_1=\sum_{n=1}^NP_n(\xi)(t-t_0)^{\alpha_n},\ g_1=\sum_{m=1}^MQ_m(\xi)(t-t_0)^{\beta_m},\ h_1=\sum_{n=1}^N\sum_{m=1}^mU_{nm}(\xi)(t-t_0)^{\alpha_n+\beta_m},\
\end{eqnarray}
\begin{eqnarray}
P'''_{n}=-\alpha_n P_n-\frac{P'_{n}}{6\Psi_1}(\xi\Psi_1-9c_1),
\end{eqnarray}
\begin{eqnarray}
Q'''_{m}=-\beta_m Q_m-\frac{Q'_{m}}{6\Psi_1}(\xi\Psi_1-9c_1),
\end{eqnarray}
\begin{eqnarray}
U'''_{nm}=-(\alpha_n+\beta_m)U_{nm}-\frac{U'_{nm}}{\Psi_1}(2\xi\Psi_1-9c_1)
-\frac{P'_{n}}2\left[6Q''_{m}-Q_m\big(\xi-3c_1\Psi^{-1}\big)\right]
+\frac{Q'_{m}}2\left[6P''_{n}-P_n\big(\xi-3c_1\Psi^{-1}\big)\right],
\end{eqnarray}
\end{subequations}
where $N$ and $M$ are arbitrary integers and $\alpha_n,\ n=1,\ 2,\ \cdots, N$ and $\beta_m,\ m=1,\ 2,\ \cdots, M$ are arbitrary constants defined on the Grassmann even algebra.

The solution of the BSKdV-2 system \eqref{bos} and then the SKdV equation \eqref{Phi} can be obtained from the nonauto-BT theorem 3,
\begin{subequations}\label{Ex2}
\begin{eqnarray}
u_0=\left(\Psi_1^2-\frac{(\sqrt{\Psi_1})_{\xi\xi}}{\sqrt{\Psi_1}}
+2\Psi_{1\xi}\tanh(w)-2\Psi_1^2\tanh^2(w)\right)(t-t_0)^{-2/3}.
\end{eqnarray}
\begin{eqnarray}
p=\left\{-\Psi_1^2f_1\mbox{sech}^2(w)-(\Psi_1f_1)_\xi \tanh(w)+\frac12[\Psi_1^{-1}(\Psi_1f_1)_\xi]_\xi
-\frac{f_1}{12\Psi_1}(\xi\Psi_1-3c_1)\right\}(t-t_0)^{-2/3}.
\end{eqnarray}
\begin{eqnarray}
q=\left\{-\Psi_1^2g_1\mbox{sech}^2(w)-(\Psi_1g_1)_\xi \tanh(w)+\frac12[\Psi_1^{-1}(\Psi_1g_1)_\xi]_\xi
-\frac{g_1}{12\Psi_1}(\xi\Psi_1-3c_1)\right\}(t-t_0)^{-2/3}.
\end{eqnarray}
\begin{eqnarray}
u_1&=&\Psi_1^3\left(\frac{h_1}{t-t_0}-\lambda_1\right)\tanh^3(w)
+\frac{\Psi_1}2\left(3\lambda_1\Psi_1\Psi_{1\xi}
-\frac{3h_1\Psi_{1\xi}+2\Psi_1h_{1\xi}}{t-t_0}\right)\tanh^2(w)
+\left(\frac{h_1}{12}\frac{9\Psi_{1\xi}^2
+2\Psi_1(\xi\Psi_\xi-3c_1)}{t-t_0}\right.
\nonumber\\
&&+\frac{2\Psi_{1\xi}h_{1\xi}+\Psi_1h_{1\xi\xi}}{2(t-t_0)}
\left.-\frac{\lambda_1}{12\Psi_1}[9\Psi_{1\xi}^2
+2\Psi_1(\xi\Psi_1-3c_1)]\right)\tanh(w)
+\frac{\lambda_1}{24\Psi_1^3}(2\Psi_1^3-3c_1\Psi_1\Psi_{1\xi}
+2\xi\Psi_1^2\Psi_{1\xi}+3\Psi_{1\xi}^3)\nonumber\\
&&+\frac{g_1}{8}
\left(\frac{f_{1\xi}(2\xi\Psi_1-3c_1)}{3\Psi_1(t-t_0)}-f_{1t}\right)
+\frac{f_1}{8}
\left(g_{1t}-\frac{g_{1\xi}(2\xi\Psi_1-3c_1)}{3\Psi_1(t-t_0)}\right)
-\frac{h_1(3\Psi_{1\xi}^3+\Psi_1^3-3c_1\Psi_1\Psi_{1\xi}
+2\xi\Psi_{1\xi}\Psi_1^2)}{24\Psi_{1}^3(t-t_0)}
\nonumber\\
&&+\frac{2(g_{1\xi\xi}f_{1\xi}-f_{1\xi\xi}g_{1\xi})\Psi_1-\Psi_{1\xi}h_{1\xi\xi}}
{4\Psi_1(t-t_0)}-\frac{h_{1\xi}}{8}\left(\frac{2\Psi_{1\xi}^2+c_1\Psi_1}
{\Psi_1^2(t-t_0)}-1\right).
\end{eqnarray}
\end{subequations}
It is clear that the symmetry reduction \eqref{g1} of the last section is only equivalent to a special case of \eqref{Ex2} with \eqref{rfgh1} and $N=M=1,$ $ \alpha_1=\beta_1$.

When $N=M=1,$ $\alpha_1=\beta_1,\ h_1=0$ and $f_1$ and $g_1$ are constants defined on $G_e$, the solution \eqref{Ex2} becomes much simpler, especially for the field $u_1$,
\begin{subequations}\label{Ex21}
\begin{eqnarray}
u_0=\left(\Psi_1^2-\frac{(\sqrt{\Psi_1})''}{\sqrt{\Psi_1}}
+2\Psi'_{1}\tanh(w)-2\Psi_1^2\tanh^2(w)\right)(t-t_0)^{-2/3},
\end{eqnarray}
\begin{eqnarray}
p=f_1\left[-\Psi_1^2\mbox{sech}^2(w)-\Psi'_{1} \tanh(w)+\frac12(\Psi_1^{-1}\Psi'_{1})'
-\frac1{12\Psi_1}(\xi\Psi_1-3c_1)\right](t-t_0)^{-2/3},
\end{eqnarray}
\begin{eqnarray}
q=g_1\left[-\Psi_1^2\mbox{sech}^2(w)-\Psi'_{1} \tanh(w)+\frac12(\Psi_1^{-1}\Psi'_{1})'
-\frac1{12\Psi_1}(\xi\Psi_1-3c_1)\right](t-t_0)^{-2/3},
\end{eqnarray}
\begin{eqnarray}
u_1&=&\lambda_1\left\{-\Psi_1^3\tanh^3(w)
+\frac32\Psi_1^2\Psi'_{1}\tanh^2(w)
-\frac1{12\Psi_1}\left[9\Psi'^2_{1}
+2\Psi_1(\xi\Psi_1-3c_1)\right]\tanh(w)\right.\nonumber\\
&&
\left.+\frac1{24\Psi_1^3}\left(2\Psi_1^3-3c_1\Psi_1\Psi'_{1}
+2\xi\Psi_1^2\Psi'_{1}+3\Psi'^3_{1}\right)\right\}.
\end{eqnarray}
\end{subequations}
\bf Example 3 \it Soliton-Cnoidal waves. \rm In nonlinear systems, the solitons (or solitary waves) and the periodic cnoidal waves are two types of typical excitations. To find the interaction solutions between solitons and cnoidal periodic waves is quite difficult. However, it is quite simple to find the soliton-cnoidal interaction solutions by using the method proposed in this section, because one of the solitons has been straightened for the $w$ field. Thus we can look for the solutions with one straight line ($k_1k_0x-\omega_0$) plus an undetermined traveling wave ($\Psi(k_1x-\omega_0t)$) for the $w$ field
\begin{eqnarray}\label{cs}
w=k_1k_0x-\omega_0t+\Psi(\eta),\qquad \eta\equiv k_1x-\omega_0t.
\end{eqnarray}
Substituting \eqref{cs} into the nonauto-BT, we have
\begin{subequations}\label{fgh2}
\begin{eqnarray}\label{eqw2}
\Psi_1\Psi''_{1}-\frac32\Psi'^2-2\Psi_1^4
+\frac12\gamma_2\Psi_1^2+\gamma_1\Psi_1, \quad \omega_1=k_1\lambda-\frac12k_1^3\gamma_2,\ \omega_0=-k_1^3\gamma_1+k_1k_0\lambda-\frac12\gamma_2k_0k_1^3,
\end{eqnarray}
\begin{eqnarray}\label{eqf2}
\Psi_1\left(k_1^{-3}f_{2t}+f_{2\eta\eta\eta}\right)
-\frac14f_{2\eta}\left[\left(\gamma_2+2\lambda k_1^{-2}\right)\Psi_1-6\gamma_1\right]=0, f\equiv f_2(\eta,\ t), \ g\equiv g_2(\eta,\ t),
\end{eqnarray}
\begin{eqnarray}\label{eqg2}
\Psi_1\left(k_1^{-3}g_{2t}+g_{2\eta\eta\eta}\right)
-\frac14g_{2\eta}\left[\left(\gamma_2+2\lambda k_1^{-2}\right)\Psi_1-6\gamma_1\right]=0,
\ h\equiv h_2(\eta,\ t)-\lambda_1 t,
\end{eqnarray}
\begin{eqnarray}\label{equ2}
&&\Psi_1(k_1^{-3}h_{2t}+h_{2\eta\eta\eta})-h_{2\eta}(\gamma\Psi_1+3\gamma)
+\frac14f_{2\eta}\left[12\Psi_1g_{2\eta\eta}+g_2\left((2\lambda k_1^{-2}-3\gamma_2)\Psi_1-6\gamma_1\right)\right]\nonumber\\
&&\quad
+\frac14g_{2\eta}\left[12\Psi_1f_{2\eta\eta}+f_2\left((2\lambda k_1^{-2}-3\gamma_2)\Psi_1-6\gamma_1\right)\right]=0,
\end{eqnarray}
\end{subequations}
with arbitrary constants $\lambda,\ k_1,\ k_0,\ \gamma_1$ and $\gamma_2$ defined on the $G_e$ algebra.

The general solution of \eqref{eqw2} reads
\begin{eqnarray}\label{solw2}
\int \frac{\mbox{d} \Psi_1}{W(\Psi_1)}=\eta-\eta_0, \quad W\equiv W(\Psi_1)=\sqrt{4\Psi_1^4+\gamma_3\Psi_1^3+\gamma_2\Psi_1^2+\gamma_1\Psi_1}
\end{eqnarray}
with two further arbitrary constants $\gamma_3$ and $\eta_0$.

If the parameters $\gamma_1,\ \gamma_2,$ and $\gamma_3$ are rewritten as
\begin{eqnarray}\label{g123}
\gamma_1=4w_0\mu^3\nu,\quad
\gamma_2=4\mu^2(3\nu-1-m^2),\quad
\gamma_3=-\frac{4\mu[2\nu(1+m^2)-3\nu^2-m^2]}{\nu w_0},
\quad w_0=\pm \sqrt{\frac{(m^2-\nu)(1-\nu)}{\nu}},
\end{eqnarray}
with arbitrary constants $\mu,\ \nu$ and $m$ defined on $G_e$, the solution \eqref{solw2} can be expressed by the Jaccobi elliptic functions, say,
\begin{eqnarray}\label{solw21}
\Psi_1=\frac{\mu w_0}{1-\nu \mbox{sn}^2(\mu \eta,\ m)}-w_0\mu, \qquad (k_0=-w_0\mu),
\end{eqnarray}
and then
\begin{eqnarray}\label{solw22}
\Psi=w_0E_\pi(\mbox{sn}^2(\mu\eta,\ m),\ \nu,\ m), \
\end{eqnarray}
where the function $E_\pi(z,\ \nu,\ m)$ is the third kind of incomplete elliptic integral
\begin{eqnarray}\label{solw22}
E_\pi(z,\ \nu,\ m)=\int_0^z\frac{\mbox{d}t}{(1-\nu t^2)\sqrt{(1-t)(1-m^2 t)}}.
\end{eqnarray}

Some types of special solutions of \eqref{fgh2} for the fields $f_2,\ g_2$ and $h_2$ can be obtained by the variable separation approach and the linear superpositions:
\begin{subequations}\label{rfgh2}
\begin{eqnarray}
f_2=\sum_{n=1}^NP_n(\Psi_1)e^{\alpha_n t},\
g_2=\sum_{m=1}^MQ_m(\Psi_1)e^{\beta_m t},\ h_2=\sum_{n=1}^N\sum_{m=1}^mU_{nm}(\Psi_1)e^{(\alpha_n+\beta_m)t},\
\end{eqnarray}
\begin{eqnarray}
\Psi_1W^2P'''_{n}+3\Psi_1WW'P''_{n}
+\left(24\Psi_1^3+3\gamma_3\Psi_1^2+\frac34\gamma_2\Psi_1
-\frac{\lambda\Psi_1}{2k_1}-\frac32\gamma_1\right)P'_{n}+\frac{\alpha_n \Psi_1 P_n}{k_1^3W},
\end{eqnarray}
\begin{eqnarray}
\Psi_1W^2Q'''_{m}+3\Psi_1WW'Q''_{m}
+\left(24\Psi_1^3+3\gamma_3\Psi_1^2+\frac34\gamma_2\Psi_1
-\frac{\lambda\Psi_1}{2k_1}-\frac32\gamma_1\right)Q'_{m}+\frac{\beta_m \Psi_1 Q_m}{k_1^3W},
\end{eqnarray}
\begin{eqnarray}
&&\Psi_1W^2U'''_{mn}+3\Psi_1WW'U''_{mn}
+\left[\Psi_1(W'^2+WW''-\gamma_2)-3\gamma_1\right]U'_{mn}
+\frac{(\alpha_n+\beta_m) \Psi_1 U_{mn}}{k_1^3W}
\nonumber\\
&&\quad +P'\left[3\Psi_1W^2Q''+\left(\frac{\lambda\Psi_1}{2k_1^2}-\frac32\gamma_1
-\frac34\gamma_2\Psi_1\right)Q\right]
+Q'\left[3\Psi_1W^2P''+\left(\frac{\lambda\Psi_1}{2k_1^2}-\frac32\gamma_1
-\frac34\gamma_2\Psi_1\right)P\right]=0,
\end{eqnarray}
\end{subequations}
with new arbitrary constants $\alpha_n,\ n=1,\ 2,\ \cdots,\ N$ and $\beta_m,\ m=1,\ 2,\ \cdots,\ M$ defined on $G_e$.

The corresponding solution of the BSKdV-2 system reads
\begin{subequations}\label{Ex3}
\begin{eqnarray}
u_0=k_1^2(2\Psi_1^2\mbox{sech}^2(w)+2W\tanh(w)-\Psi_1^2)+\frac{\lambda}6
-\frac{k_1^2W}{4\Psi_1^2}(2\Psi_1W'-W),
\end{eqnarray}
\begin{eqnarray}
p=f_2k_1^2\Psi_1^2\tanh^2(w)-k_1^2(f_2\Psi_1)_\eta \tanh(w)+\frac{f_2}{24\Psi_1^2}\left[6k_1^2W^2-6\Psi_1k_1^2\gamma_1
-\Psi_1^2(2\lambda-3k_1^2\gamma_2)\right]
+\frac{k_1^2}{2\Psi_1}(f_{2\eta}\Psi_1)_\eta,
\end{eqnarray}
\begin{eqnarray}
q=g_2k_1^2\Psi_1^2\tanh^2(w)-k_1^2(g_2\Psi_1)_\eta \tanh(w)+\frac{g_2}{24\Psi_1^2}\left[6k_1^2W^2-6\Psi_1k_1^2\gamma_1
-\Psi_1^2(2\lambda-3k_1^2\gamma_2)\right]
+\frac{k_1^2}{2\Psi_1}(g_{2\eta}\Psi_1)_\eta,
\end{eqnarray}
\begin{eqnarray}
u_1&=&k_1^3(h_2-\lambda_1 t)\Psi_1^3\tanh^3(w)
-\frac{k_1^3}2\Psi_1[2h_\eta \Psi_{1}+3W(h_2-\lambda_1 t)]\tanh^2(w)
+k_1^3\left[\frac14\left(\lambda_1\gamma_2t+2h_{2\eta\eta}\right)\Psi_1-\frac{3\lambda_1}{4\Psi_1}W^2t+Wh_{2\eta}\right.
\nonumber\\
&&\left.+\frac12\lambda_1\gamma_1t-\frac14\left(2\gamma_1+\gamma_2\Psi_1
-\frac{3W^2}{\Psi_1}\right)\right]\tanh(w)
+\frac{k_1^3}2\left[g_{2\eta\eta}-\frac12\left(\frac{\lambda}{3k_1^2}
-\frac{\gamma_2}2-\frac{\gamma_1}{2\Psi_1}\right)g_2\right]f_{2\eta}
-\frac{k_1^3}2\left[f_{2\eta\eta}-\frac12\left(\frac{\lambda}{3k_1^2}
-\frac{\gamma_2}2\right.\right.\nonumber\\
&&
\left.\left.-\frac{\gamma_1}{2\Psi_1}\right)f_2\right]g_{2\eta}
+\frac{f_2g_{2t}-g_2f_{2t}}{8}+\frac{k_1^3 W h_2}{8\Psi_3}(\gamma_1\Psi_1
-W^2+\gamma_2\Psi_1^2)+\frac{W^3\lambda_1t}{8\Psi_1^3}
-\frac{k_1^3}{8\Psi_1}\left[\gamma_1h_{2\eta}
-W(2h_{2\eta\eta}-\gamma_2\lambda_1t)\right]\nonumber\\
&&
-\frac{k_1^3W}{8\Psi_1^2}(2Wh_{2\eta}+\gamma_1\lambda_1t)
+\frac{\lambda_1+3h_{2t}}{24}.
\end{eqnarray}
\end{subequations}
It is clear that if we take $N=M=1,\ \beta_1=\alpha_1,\ \lambda_1=0,$ $f_2, \ g_2,\ h_2, $ and $\alpha_2$ are constants defined on $G_e$, then the solution \eqref{Ex3} is reduced back to the special symmetry reductions \eqref{g2}--\eqref{Phi1} discussed in the last section.

\section{Conclusions}

In summary, the simple bosonization approach proposed in the previous Letter \cite{Boson} is extended such that all the bosonized fields are defined on even Grassmann algebra $G_e$ \eqref{Ge}. With $N$ fermionic parameters,
the bosonization procedure of the supersymmetric systems has been
successfully applied to the SKdV equation in detail. Such an
integrable nonlinear system is simplified to the KdV equation
together with several linear differential equations. Though the supersymmetric KdV is bosonized to BSKdV-$N$ systems for arbitrary $N$, all the boson fields are different to the traditional non-supersymmetric boson systems, because they are defined on $G_e$ algebra.

The BSKdV-2 system is proved to be Painlev\'e integrable, which means that the model possesses Painlev\'e property. Starting from the standard truncated Painlev\'e expansion, a B\"acklund transformation with a free spectral parameter is found. Furthermore, we find that the residual of the BT, i.e., the truncated Painlev\'e expansion, is a nonlocal symmetry of the BSKdV-2 and the symmetry is defined as RS (residual symmetry). It is proved that the RS is just equivalent to the infinitesimal form, the generator, of the BT. For sake of the free spectral parameter in RS, infinitely many nonlocal symmetries can be found in two ways. Especially, the higher order nonlocal symmetries can be obtained simply by differentiating the RS any times with respect to the spectral parameter.

The nonlocal symmetry RS is nonlocal for the original BSKdV-2 system, however, it can be successfully localized by enlarging it to the prolonged BSKdV-2 system \eqref{bose}. Thanks to the localization processing, the nonlocal symmetry RS is used to find possible symmetry reductions. Especially, the usual Painlev\'e II reductions and the cnoidal wave reduction solutions can be extended an additional soliton. This fact implies that starting from any seed solution of an integrable model, the BT will introduce an additional soliton to the original seed, because the RS is just the infinitesimal form of BT. To manifest the correctness of this conjecture, a much simper method, the generalized tanh function expansion method, is proposed.
The generalized tanh function expansion method for the BSKdV-2 system leads to a nonauto-BT theorem, which strengthens a single soliton
to a straight-line solution. Hereafter, to add a soliton to any seed wave becomes a quite simple work: simply plus a straight line solution on a general solution in the nonauto-BT leads to the interaction solution between soliton and arbitrary other seed waves for the original BSKdV-2 system. Using the nonauto-BT theorem, various exact explicit solutions of the BSKdV-2 system are obtained.

In this paper, we only investigate the properties and exact solutions of the BSKdV-2 system, nonetheless, all the results are similar for arbitrary BSKdV-$N$ systems. For instance, one can prove that the BSKdV-$N$ system \eqref{Eqn} is Painlev\'e integrable because of the existence of the Painlev\'e expansion
\begin{eqnarray}
u_0=\sum_{j=0}^{\infty}u_{0i}\phi^{j-2},\
v_{i_{1}i_{2} \cdots i_{2n-1}}=\sum_{j=0}^{\infty}v_{i_{1}i_{2} \cdots
i_{2n-1},j}\phi^{j-2},\
u_{i_{1}i_{2}\cdots i_{2n}}=\sum_{j=0}^{\infty}u_{i_{1}i_{2}\cdots i_{2n},j}\phi^{j-3},
\label{ppn}
\end{eqnarray}
where the expansion coefficients
$u_{0j}$, $u_{i_{1}i_{2}
\cdots i_{2n},j}$ and $v_{i_{1}i_{2} \cdots
i_{2n-1},j}$ are determined by $3\times 2^N$ arbitrary functions $\phi, \ u_{04},\ u_{06},\ $ $ v_{i_{1}i_{2} \cdots
i_{2n-1},0},$ $ v_{i_{1}i_{2} \cdots
i_{2n-1},4},$ $ v_{i_{1}i_{2} \cdots
i_{2n-1},5},$ $ u_{i_{1}i_{2}
\cdots i_{2n},0},$ $ u_{i_{1}i_{2}
\cdots i_{2n},5},$ and $u_{i_{1}i_{2}
\cdots i_{2n},7}$.

The bosonization approach can be applied to not only the
supersymmetric integrable systems but also all the models with
fermion fields no matter whether they are integrable or not.
It should be emphasized that the solutions obtained via the
bosonization procedure are completely different from those obtained
via other methods such as the bilinear approach \cite{BL}.

The results of our previous Letter \cite{Boson} and this paper
show that for the SKdV equation there exist various kinds of
localized excitations and the interaction solution between different types of waves. For instance, the solutions \eqref{cs}-\eqref{Ex3} demonstrate that for one excitation of the usual non-supersymmetric KdV model, there exist infinitely many possible excitations for SKdV system.

The richness of the soliton structure and other types of excitations of the classical SKdV reveal some open problems in both classical and quantum
theories.
For instance, generically the fermionic fields take value on an infinite Grassmann algebra, the BSKdV-$N$ of this paper is only defined on an infinite dimensional Grassmann subalgebra, the even Grassmann algebra $G_e$. Hence, one of the important problems is how to obtain an extension
to the case of infinite full Grassmann algebra.
In the quantum level, one of the most
important topics may be how to reflect
the richness of the localized excitations in the usual quantization
procedure of the supersymmetry models\cite{Kulish}?

\subsection{Acknowledgement}
The work is sponsored by the National Natural Science Foundation of
China (No. 11175092 and No. 11275123), Shanghai Knowledge Service Platform for Trustworthy Internet of Things (No. ZF1213), and
K. C. Wong Magna Fund in Ningbo University.

\end{document}